\documentclass[longauth]{aa}

\usepackage{graphicx}
\usepackage{natbib}
\usepackage{scalerel}
\usepackage{comment}
\usepackage{multirow}

\usepackage{makecell}
\usepackage[dvipsnames]{xcolor}

\usepackage{soul}

\usepackage{txfonts}
\usepackage[pdfencoding=auto,psdextra]{hyperref}
\hypersetup{
    colorlinks=true,
    linkcolor=blue,
    filecolor=magenta,      
    urlcolor=blue,
    citecolor=blue
}
\makeatletter
\renewcommand*\aa@pageof{, page \thepage{} of \pageref*{LastPage}}
\makeatother

\usepackage[utf8]{inputenc}

\usepackage[switch, modulo]{lineno}

\usepackage{euclid}

\newcommand{\GCsp}{\text{GC}\ensuremath{_\text{sp}}}
\newcommand{\GCph}{\text{GC}\ensuremath{_\text{ph}}}
\newcommand{\WL}{\text{WL}}
\newcommand{\XC}{\text{XC}}
\newcommand{\Omegam}{\Omega_{\mathrm{m},0}}
\newcommand{\Omegab}{\Omega_{\mathrm{b},0}}

\defcitealias{IST:paper1}{EP:VII}\usepackage[nameinlink,noabbrev]{cleveref}

\newenvironment{mytable}[1][ht]{\begin{table}[#1]\centering\small}{\end{table}} 

\begin{document}
%
%
   \title{\Euclid\/ preparation.  Improving  cosmological constraints using 
   a new multi-tracer method with the 
   spectroscopic and photometric samples
   }

\newcommand{\orcid}[1]{} 
\author{Euclid Collaboration: F.~Dournac\inst{\ref{aff1}}
\and A.~Blanchard\orcid{0000-0001-8555-9003}\thanks{\email{alain.blanchard@irap.omp.eu}}\inst{\ref{aff1}}
\and S.~Ili\'c\orcid{0000-0003-4285-9086}\inst{\ref{aff2},\ref{aff1}}
\and B.~Lamine\orcid{0000-0002-9416-2320}\inst{\ref{aff1}}
\and I.~Tutusaus\orcid{0000-0002-3199-0399}\inst{\ref{aff1}}
\and A.~Amara\inst{\ref{aff3}}
\and S.~Andreon\orcid{0000-0002-2041-8784}\inst{\ref{aff4}}
\and N.~Auricchio\orcid{0000-0003-4444-8651}\inst{\ref{aff5}}
\and H.~Aussel\orcid{0000-0002-1371-5705}\inst{\ref{aff6}}
\and M.~Baldi\orcid{0000-0003-4145-1943}\inst{\ref{aff7},\ref{aff5},\ref{aff8}}
\and S.~Bardelli\orcid{0000-0002-8900-0298}\inst{\ref{aff5}}
\and C.~Bodendorf\inst{\ref{aff9}}
\and D.~Bonino\orcid{0000-0002-3336-9977}\inst{\ref{aff10}}
\and E.~Branchini\orcid{0000-0002-0808-6908}\inst{\ref{aff11},\ref{aff12},\ref{aff4}}
\and S.~Brau-Nogue\inst{\ref{aff1}}
\and M.~Brescia\orcid{0000-0001-9506-5680}\inst{\ref{aff13},\ref{aff14},\ref{aff15}}
\and J.~Brinchmann\orcid{0000-0003-4359-8797}\inst{\ref{aff16}}
\and S.~Camera\orcid{0000-0003-3399-3574}\inst{\ref{aff17},\ref{aff18},\ref{aff10}}
\and V.~Capobianco\orcid{0000-0002-3309-7692}\inst{\ref{aff10}}
\and J.~Carretero\orcid{0000-0002-3130-0204}\inst{\ref{aff19},\ref{aff20}}
\and S.~Casas\orcid{0000-0002-4751-5138}\inst{\ref{aff21}}
\and M.~Castellano\orcid{0000-0001-9875-8263}\inst{\ref{aff22}}
\and S.~Cavuoti\orcid{0000-0002-3787-4196}\inst{\ref{aff14},\ref{aff15}}
\and A.~Cimatti\inst{\ref{aff23}}
\and G.~Congedo\orcid{0000-0003-2508-0046}\inst{\ref{aff24}}
\and C.~J.~Conselice\orcid{0000-0003-1949-7638}\inst{\ref{aff25}}
\and L.~Conversi\orcid{0000-0002-6710-8476}\inst{\ref{aff26},\ref{aff27}}
\and Y.~Copin\orcid{0000-0002-5317-7518}\inst{\ref{aff28}}
\and F.~Courbin\orcid{0000-0003-0758-6510}\inst{\ref{aff29}}
\and H.~M.~Courtois\orcid{0000-0003-0509-1776}\inst{\ref{aff30}}
\and A.~Da~Silva\orcid{0000-0002-6385-1609}\inst{\ref{aff31},\ref{aff32}}
\and H.~Degaudenzi\orcid{0000-0002-5887-6799}\inst{\ref{aff33}}
\and A.~M.~Di~Giorgio\orcid{0000-0002-4767-2360}\inst{\ref{aff34}}
\and J.~Dinis\inst{\ref{aff31},\ref{aff32}}
\and M.~Douspis\inst{\ref{aff35}}
\and F.~Dubath\orcid{0000-0002-6533-2810}\inst{\ref{aff33}}
\and X.~Dupac\inst{\ref{aff27}}
\and S.~Dusini\orcid{0000-0002-1128-0664}\inst{\ref{aff36}}
\and A.~Ealet\orcid{0000-0003-3070-014X}\inst{\ref{aff28}}
\and M.~Farina\orcid{0000-0002-3089-7846}\inst{\ref{aff34}}
\and S.~Farrens\orcid{0000-0002-9594-9387}\inst{\ref{aff6}}
\and S.~Ferriol\inst{\ref{aff28}}
\and M.~Frailis\orcid{0000-0002-7400-2135}\inst{\ref{aff37}}
\and E.~Franceschi\orcid{0000-0002-0585-6591}\inst{\ref{aff5}}
\and S.~Galeotta\orcid{0000-0002-3748-5115}\inst{\ref{aff37}}
\and W.~Gillard\orcid{0000-0003-4744-9748}\inst{\ref{aff38}}
\and B.~Gillis\orcid{0000-0002-4478-1270}\inst{\ref{aff24}}
\and C.~Giocoli\orcid{0000-0002-9590-7961}\inst{\ref{aff5},\ref{aff39}}
\and B.~R.~Granett\orcid{0000-0003-2694-9284}\inst{\ref{aff4}}
\and A.~Grazian\orcid{0000-0002-5688-0663}\inst{\ref{aff40}}
\and F.~Grupp\inst{\ref{aff9},\ref{aff41}}
\and S.~V.~H.~Haugan\orcid{0000-0001-9648-7260}\inst{\ref{aff42}}
\and W.~Holmes\inst{\ref{aff43}}
\and I.~Hook\orcid{0000-0002-2960-978X}\inst{\ref{aff44}}
\and F.~Hormuth\inst{\ref{aff45}}
\and A.~Hornstrup\orcid{0000-0002-3363-0936}\inst{\ref{aff46},\ref{aff47}}
\and P.~Hudelot\inst{\ref{aff48}}
\and K.~Jahnke\orcid{0000-0003-3804-2137}\inst{\ref{aff49}}
\and E.~Keih\"anen\orcid{0000-0003-1804-7715}\inst{\ref{aff50}}
\and S.~Kermiche\orcid{0000-0002-0302-5735}\inst{\ref{aff38}}
\and A.~Kiessling\orcid{0000-0002-2590-1273}\inst{\ref{aff43}}
\and M.~Kilbinger\orcid{0000-0001-9513-7138}\inst{\ref{aff51}}
\and B.~Kubik\orcid{0009-0006-5823-4880}\inst{\ref{aff28}}
\and M.~K\"ummel\orcid{0000-0003-2791-2117}\inst{\ref{aff41}}
\and M.~Kunz\orcid{0000-0002-3052-7394}\inst{\ref{aff52}}
\and H.~Kurki-Suonio\orcid{0000-0002-4618-3063}\inst{\ref{aff53},\ref{aff54}}
\and S.~Ligori\orcid{0000-0003-4172-4606}\inst{\ref{aff10}}
\and P.~B.~Lilje\orcid{0000-0003-4324-7794}\inst{\ref{aff42}}
\and V.~Lindholm\orcid{0000-0003-2317-5471}\inst{\ref{aff53},\ref{aff54}}
\and I.~Lloro\inst{\ref{aff55}}
\and D.~Maino\inst{\ref{aff56},\ref{aff57},\ref{aff58}}
\and E.~Maiorano\orcid{0000-0003-2593-4355}\inst{\ref{aff5}}
\and O.~Mansutti\orcid{0000-0001-5758-4658}\inst{\ref{aff37}}
\and O.~Marggraf\orcid{0000-0001-7242-3852}\inst{\ref{aff59}}
\and K.~Markovic\orcid{0000-0001-6764-073X}\inst{\ref{aff43}}
\and N.~Martinet\orcid{0000-0003-2786-7790}\inst{\ref{aff60}}
\and F.~Marulli\orcid{0000-0002-8850-0303}\inst{\ref{aff61},\ref{aff5},\ref{aff8}}
\and R.~Massey\orcid{0000-0002-6085-3780}\inst{\ref{aff62}}
\and S.~Maurogordato\inst{\ref{aff63}}
\and E.~Medinaceli\orcid{0000-0002-4040-7783}\inst{\ref{aff5}}
\and S.~Mei\orcid{0000-0002-2849-559X}\inst{\ref{aff64}}
\and Y.~Mellier\inst{\ref{aff65},\ref{aff48}}
\and M.~Meneghetti\orcid{0000-0003-1225-7084}\inst{\ref{aff5},\ref{aff8}}
\and E.~Merlin\orcid{0000-0001-6870-8900}\inst{\ref{aff22}}
\and G.~Meylan\inst{\ref{aff29}}
\and M.~Moresco\orcid{0000-0002-7616-7136}\inst{\ref{aff61},\ref{aff5}}
\and L.~Moscardini\orcid{0000-0002-3473-6716}\inst{\ref{aff61},\ref{aff5},\ref{aff8}}
\and E.~Munari\orcid{0000-0002-1751-5946}\inst{\ref{aff37}}
\and S.-M.~Niemi\inst{\ref{aff66}}
\and J.~W.~Nightingale\orcid{0000-0002-8987-7401}\inst{\ref{aff67},\ref{aff68}}
\and C.~Padilla\orcid{0000-0001-7951-0166}\inst{\ref{aff19}}
\and S.~Paltani\orcid{0000-0002-8108-9179}\inst{\ref{aff33}}
\and F.~Pasian\orcid{0000-0002-4869-3227}\inst{\ref{aff37}}
\and K.~Pedersen\inst{\ref{aff69}}
\and W.~J.~Percival\orcid{0000-0002-0644-5727}\inst{\ref{aff70},\ref{aff71},\ref{aff72}}
\and V.~Pettorino\inst{\ref{aff66}}
\and S.~Pires\orcid{0000-0002-0249-2104}\inst{\ref{aff6}}
\and G.~Polenta\orcid{0000-0003-4067-9196}\inst{\ref{aff73}}
\and M.~Poncet\inst{\ref{aff74}}
\and L.~A.~Popa\inst{\ref{aff75}}
\and L.~Pozzetti\orcid{0000-0001-7085-0412}\inst{\ref{aff5}}
\and F.~Raison\orcid{0000-0002-7819-6918}\inst{\ref{aff9}}
\and R.~Rebolo\inst{\ref{aff76},\ref{aff77}}
\and A.~Renzi\orcid{0000-0001-9856-1970}\inst{\ref{aff78},\ref{aff36}}
\and J.~Rhodes\inst{\ref{aff43}}
\and G.~Riccio\inst{\ref{aff14}}
\and E.~Romelli\orcid{0000-0003-3069-9222}\inst{\ref{aff37}}
\and M.~Roncarelli\orcid{0000-0001-9587-7822}\inst{\ref{aff5}}
\and E.~Rossetti\inst{\ref{aff7}}
\and R.~Saglia\orcid{0000-0003-0378-7032}\inst{\ref{aff41},\ref{aff9}}
\and D.~Sapone\orcid{0000-0001-7089-4503}\inst{\ref{aff79}}
\and P.~Schneider\orcid{0000-0001-8561-2679}\inst{\ref{aff59}}
\and A.~Secroun\orcid{0000-0003-0505-3710}\inst{\ref{aff38}}
\and G.~Seidel\orcid{0000-0003-2907-353X}\inst{\ref{aff49}}
\and M.~Seiffert\orcid{0000-0002-7536-9393}\inst{\ref{aff43}}
\and S.~Serrano\orcid{0000-0002-0211-2861}\inst{\ref{aff80},\ref{aff81},\ref{aff82}}
\and C.~Sirignano\orcid{0000-0002-0995-7146}\inst{\ref{aff78},\ref{aff36}}
\and G.~Sirri\orcid{0000-0003-2626-2853}\inst{\ref{aff8}}
\and L.~Stanco\orcid{0000-0002-9706-5104}\inst{\ref{aff36}}
\and C.~Surace\orcid{0000-0003-2592-0113}\inst{\ref{aff60}}
\and P.~Tallada-Cresp\'{i}\orcid{0000-0002-1336-8328}\inst{\ref{aff83},\ref{aff20}}
\and D.~Tavagnacco\orcid{0000-0001-7475-9894}\inst{\ref{aff37}}
\and A.~N.~Taylor\inst{\ref{aff24}}
\and I.~Tereno\inst{\ref{aff31},\ref{aff84}}
\and R.~Toledo-Moreo\orcid{0000-0002-2997-4859}\inst{\ref{aff85}}
\and F.~Torradeflot\orcid{0000-0003-1160-1517}\inst{\ref{aff20},\ref{aff83}}
\and E.~A.~Valentijn\inst{\ref{aff86}}
\and L.~Valenziano\orcid{0000-0002-1170-0104}\inst{\ref{aff5},\ref{aff87}}
\and T.~Vassallo\orcid{0000-0001-6512-6358}\inst{\ref{aff41},\ref{aff37}}
\and A.~Veropalumbo\orcid{0000-0003-2387-1194}\inst{\ref{aff4},\ref{aff12}}
\and Y.~Wang\orcid{0000-0002-4749-2984}\inst{\ref{aff88}}
\and A.~Zacchei\orcid{0000-0003-0396-1192}\inst{\ref{aff37},\ref{aff89}}
\and G.~Zamorani\orcid{0000-0002-2318-301X}\inst{\ref{aff5}}
\and J.~Zoubian\inst{\ref{aff38}}
\and E.~Zucca\orcid{0000-0002-5845-8132}\inst{\ref{aff5}}
\and A.~Biviano\orcid{0000-0002-0857-0732}\inst{\ref{aff37},\ref{aff89}}
\and M.~Bolzonella\orcid{0000-0003-3278-4607}\inst{\ref{aff5}}
\and A.~Boucaud\orcid{0000-0001-7387-2633}\inst{\ref{aff64}}
\and E.~Bozzo\orcid{0000-0002-8201-1525}\inst{\ref{aff33}}
\and C.~Burigana\orcid{0000-0002-3005-5796}\inst{\ref{aff90},\ref{aff87}}
\and C.~Colodro-Conde\inst{\ref{aff76}}
\and G.~De~Lucia\orcid{0000-0002-6220-9104}\inst{\ref{aff37}}
\and D.~Di~Ferdinando\inst{\ref{aff8}}
\and J.~A.~Escartin~Vigo\inst{\ref{aff9}}
\and R.~Farinelli\inst{\ref{aff5}}
\and J.~Gracia-Carpio\inst{\ref{aff9}}
\and G.~Mainetti\inst{\ref{aff91}}
\and M.~Martinelli\orcid{0000-0002-6943-7732}\inst{\ref{aff22},\ref{aff92}}
\and N.~Mauri\orcid{0000-0001-8196-1548}\inst{\ref{aff23},\ref{aff8}}
\and C.~Neissner\orcid{0000-0001-8524-4968}\inst{\ref{aff19},\ref{aff20}}
\and Z.~Sakr\orcid{0000-0002-4823-3757}\inst{\ref{aff93},\ref{aff1},\ref{aff94}}
\and V.~Scottez\inst{\ref{aff65},\ref{aff95}}
\and M.~Tenti\orcid{0000-0002-4254-5901}\inst{\ref{aff8}}
\and M.~Viel\orcid{0000-0002-2642-5707}\inst{\ref{aff89},\ref{aff37},\ref{aff96},\ref{aff97},\ref{aff98}}
\and M.~Wiesmann\orcid{0009-0000-8199-5860}\inst{\ref{aff42}}
\and Y.~Akrami\orcid{0000-0002-2407-7956}\inst{\ref{aff99},\ref{aff100}}
\and V.~Allevato\orcid{0000-0001-7232-5152}\inst{\ref{aff14}}
\and S.~Anselmi\orcid{0000-0002-3579-9583}\inst{\ref{aff36},\ref{aff78},\ref{aff101}}
\and C.~Baccigalupi\orcid{0000-0002-8211-1630}\inst{\ref{aff96},\ref{aff37},\ref{aff97},\ref{aff89}}
\and A.~Balaguera-Antolinez\orcid{0000-0001-5028-3035}\inst{\ref{aff76},\ref{aff77}}
\and M.~Ballardini\orcid{0000-0003-4481-3559}\inst{\ref{aff102},\ref{aff5},\ref{aff103}}
\and L.~Blot\orcid{0000-0002-9622-7167}\inst{\ref{aff104},\ref{aff101}}
\and S.~Borgani\orcid{0000-0001-6151-6439}\inst{\ref{aff105},\ref{aff89},\ref{aff37},\ref{aff97}}
\and S.~Bruton\orcid{0000-0002-6503-5218}\inst{\ref{aff106}}
\and R.~Cabanac\orcid{0000-0001-6679-2600}\inst{\ref{aff1}}
\and A.~Calabro\orcid{0000-0003-2536-1614}\inst{\ref{aff22}}
\and G.~Canas-Herrera\orcid{0000-0003-2796-2149}\inst{\ref{aff66},\ref{aff107}}
\and A.~Cappi\inst{\ref{aff5},\ref{aff63}}
\and C.~S.~Carvalho\inst{\ref{aff84}}
\and G.~Castignani\orcid{0000-0001-6831-0687}\inst{\ref{aff5}}
\and T.~Castro\orcid{0000-0002-6292-3228}\inst{\ref{aff37},\ref{aff97},\ref{aff89},\ref{aff98}}
\and K.~C.~Chambers\orcid{0000-0001-6965-7789}\inst{\ref{aff108}}
\and S.~Contarini\orcid{0000-0002-9843-723X}\inst{\ref{aff9},\ref{aff61}}
\and A.~R.~Cooray\orcid{0000-0002-3892-0190}\inst{\ref{aff109}}
\and J.~Coupon\inst{\ref{aff33}}
\and S.~Davini\orcid{0000-0003-3269-1718}\inst{\ref{aff12}}
\and B.~De~Caro\inst{\ref{aff36},\ref{aff78}}
\and S.~de~la~Torre\inst{\ref{aff60}}
\and G.~Desprez\inst{\ref{aff110}}
\and A.~D\'iaz-S\'anchez\orcid{0000-0003-0748-4768}\inst{\ref{aff111}}
\and S.~Di~Domizio\orcid{0000-0003-2863-5895}\inst{\ref{aff11},\ref{aff12}}
\and H.~Dole\orcid{0000-0002-9767-3839}\inst{\ref{aff35}}
\and S.~Escoffier\orcid{0000-0002-2847-7498}\inst{\ref{aff38}}
\and A.~G.~Ferrari\orcid{0009-0005-5266-4110}\inst{\ref{aff23},\ref{aff8}}
\and P.~G.~Ferreira\inst{\ref{aff112}}
\and I.~Ferrero\orcid{0000-0002-1295-1132}\inst{\ref{aff42}}
\and F.~Finelli\orcid{0000-0002-6694-3269}\inst{\ref{aff5},\ref{aff87}}
\and L.~Gabarra\orcid{0000-0002-8486-8856}\inst{\ref{aff112}}
\and K.~Ganga\orcid{0000-0001-8159-8208}\inst{\ref{aff64}}
\and J.~Garc\'ia-Bellido\orcid{0000-0002-9370-8360}\inst{\ref{aff99}}
\and E.~Gaztanaga\orcid{0000-0001-9632-0815}\inst{\ref{aff81},\ref{aff80},\ref{aff113}}
\and F.~Giacomini\orcid{0000-0002-3129-2814}\inst{\ref{aff8}}
\and G.~Gozaliasl\orcid{0000-0002-0236-919X}\inst{\ref{aff114},\ref{aff53}}
\and H.~Hildebrandt\orcid{0000-0002-9814-3338}\inst{\ref{aff115}}
\and A.~Jimenez~Munoz\orcid{0009-0004-5252-185X}\inst{\ref{aff116}}
\and J.~J.~E.~Kajava\orcid{0000-0002-3010-8333}\inst{\ref{aff117},\ref{aff118}}
\and V.~Kansal\orcid{0000-0002-4008-6078}\inst{\ref{aff119},\ref{aff120}}
\and D.~Karagiannis\orcid{0000-0002-4927-0816}\inst{\ref{aff121},\ref{aff122}}
\and C.~C.~Kirkpatrick\inst{\ref{aff50}}
\and L.~Legrand\orcid{0000-0003-0610-5252}\inst{\ref{aff123}}
\and G.~Libet\inst{\ref{aff74}}
\and A.~Loureiro\orcid{0000-0002-4371-0876}\inst{\ref{aff124},\ref{aff125}}
\and J.~Macias-Perez\orcid{0000-0002-5385-2763}\inst{\ref{aff116}}
\and G.~Maggio\orcid{0000-0003-4020-4836}\inst{\ref{aff37}}
\and M.~Magliocchetti\orcid{0000-0001-9158-4838}\inst{\ref{aff34}}
\and F.~Mannucci\orcid{0000-0002-4803-2381}\inst{\ref{aff126}}
\and R.~Maoli\orcid{0000-0002-6065-3025}\inst{\ref{aff127},\ref{aff22}}
\and C.~J.~A.~P.~Martins\orcid{0000-0002-4886-9261}\inst{\ref{aff128},\ref{aff16}}
\and S.~Matthew\inst{\ref{aff24}}
\and L.~Maurin\orcid{0000-0002-8406-0857}\inst{\ref{aff35}}
\and R.~B.~Metcalf\orcid{0000-0003-3167-2574}\inst{\ref{aff61},\ref{aff5}}
\and M.~Migliaccio\inst{\ref{aff129},\ref{aff130}}
\and P.~Monaco\orcid{0000-0003-2083-7564}\inst{\ref{aff105},\ref{aff37},\ref{aff97},\ref{aff89}}
\and C.~Moretti\orcid{0000-0003-3314-8936}\inst{\ref{aff96},\ref{aff98},\ref{aff37},\ref{aff89},\ref{aff97}}
\and G.~Morgante\inst{\ref{aff5}}
\and S.~Nadathur\orcid{0000-0001-9070-3102}\inst{\ref{aff113}}
\and Nicholas~A.~Walton\orcid{0000-0003-3983-8778}\inst{\ref{aff131}}
\and L.~Patrizii\inst{\ref{aff8}}
\and A.~Pezzotta\orcid{0000-0003-0726-2268}\inst{\ref{aff9}}
\and M.~P\"ontinen\orcid{0000-0001-5442-2530}\inst{\ref{aff53}}
\and V.~Popa\inst{\ref{aff75}}
\and C.~Porciani\orcid{0000-0002-7797-2508}\inst{\ref{aff59}}
\and D.~Potter\orcid{0000-0002-0757-5195}\inst{\ref{aff132}}
\and I.~Risso\orcid{0000-0003-2525-7761}\inst{\ref{aff133}}
\and P.-F.~Rocci\inst{\ref{aff35}}
\and M.~Sahl\'en\orcid{0000-0003-0973-4804}\inst{\ref{aff134}}
\and A.~G.~S\'anchez\orcid{0000-0003-1198-831X}\inst{\ref{aff9}}
\and J.~A.~Schewtschenko\inst{\ref{aff24}}
\and A.~Schneider\orcid{0000-0001-7055-8104}\inst{\ref{aff132}}
\and E.~Sefusatti\orcid{0000-0003-0473-1567}\inst{\ref{aff37},\ref{aff89},\ref{aff97}}
\and M.~Sereno\orcid{0000-0003-0302-0325}\inst{\ref{aff5},\ref{aff8}}
\and J.~Steinwagner\inst{\ref{aff9}}
\and N.~Tessore\orcid{0000-0002-9696-7931}\inst{\ref{aff135}}
\and G.~Testera\inst{\ref{aff12}}
\and R.~Teyssier\orcid{0000-0001-7689-0933}\inst{\ref{aff136}}
\and S.~Toft\orcid{0000-0003-3631-7176}\inst{\ref{aff47},\ref{aff137},\ref{aff138}}
\and S.~Tosi\orcid{0000-0002-7275-9193}\inst{\ref{aff11},\ref{aff4},\ref{aff12}}
\and A.~Troja\orcid{0000-0003-0239-4595}\inst{\ref{aff78},\ref{aff36}}
\and M.~Tucci\inst{\ref{aff33}}
\and J.~Valiviita\orcid{0000-0001-6225-3693}\inst{\ref{aff53},\ref{aff54}}
\and D.~Vergani\orcid{0000-0003-0898-2216}\inst{\ref{aff5}}
\and G.~Verza\orcid{0000-0002-1886-8348}\inst{\ref{aff139},\ref{aff140}}}
										   
\institute{Institut de Recherche en Astrophysique et Plan\'etologie (IRAP), Universit\'e de Toulouse, CNRS, UPS, CNES, 14 Av. Edouard Belin, 31400 Toulouse, France\label{aff1}
\and
Universit\'e Paris-Saclay, CNRS/IN2P3, IJCLab, 91405 Orsay, France\label{aff2}
\and
School of Mathematics and Physics, University of Surrey, Guildford, Surrey, GU2 7XH, UK\label{aff3}
\and
INAF-Osservatorio Astronomico di Brera, Via Brera 28, 20122 Milano, Italy\label{aff4}
\and
INAF-Osservatorio di Astrofisica e Scienza dello Spazio di Bologna, Via Piero Gobetti 93/3, 40129 Bologna, Italy\label{aff5}
\and
Universit\'e Paris-Saclay, Universit\'e Paris Cit\'e, CEA, CNRS, AIM, 91191, Gif-sur-Yvette, France\label{aff6}
\and
Dipartimento di Fisica e Astronomia, Universit\`a di Bologna, Via Gobetti 93/2, 40129 Bologna, Italy\label{aff7}
\and
INFN-Sezione di Bologna, Viale Berti Pichat 6/2, 40127 Bologna, Italy\label{aff8}
\and
Max Planck Institute for Extraterrestrial Physics, Giessenbachstr. 1, 85748 Garching, Germany\label{aff9}
\and
INAF-Osservatorio Astrofisico di Torino, Via Osservatorio 20, 10025 Pino Torinese (TO), Italy\label{aff10}
\and
Dipartimento di Fisica, Universit\`a di Genova, Via Dodecaneso 33, 16146, Genova, Italy\label{aff11}
\and
INFN-Sezione di Genova, Via Dodecaneso 33, 16146, Genova, Italy\label{aff12}
\and
Department of Physics "E. Pancini", University Federico II, Via Cinthia 6, 80126, Napoli, Italy\label{aff13}
\and
INAF-Osservatorio Astronomico di Capodimonte, Via Moiariello 16, 80131 Napoli, Italy\label{aff14}
\and
INFN section of Naples, Via Cinthia 6, 80126, Napoli, Italy\label{aff15}
\and
Instituto de Astrof\'isica e Ci\^encias do Espa\c{c}o, Universidade do Porto, CAUP, Rua das Estrelas, PT4150-762 Porto, Portugal\label{aff16}
\and
Dipartimento di Fisica, Universit\`a degli Studi di Torino, Via P. Giuria 1, 10125 Torino, Italy\label{aff17}
\and
INFN-Sezione di Torino, Via P. Giuria 1, 10125 Torino, Italy\label{aff18}
\and
Institut de F\'{i}sica d'Altes Energies (IFAE), The Barcelona Institute of Science and Technology, Campus UAB, 08193 Bellaterra (Barcelona), Spain\label{aff19}
\and
Port d'Informaci\'{o} Cient\'{i}fica, Campus UAB, C. Albareda s/n, 08193 Bellaterra (Barcelona), Spain\label{aff20}
\and
Institute for Theoretical Particle Physics and Cosmology (TTK), RWTH Aachen University, 52056 Aachen, Germany\label{aff21}
\and
INAF-Osservatorio Astronomico di Roma, Via Frascati 33, 00078 Monteporzio Catone, Italy\label{aff22}
\and
Dipartimento di Fisica e Astronomia "Augusto Righi" - Alma Mater Studiorum Universit\`a di Bologna, Viale Berti Pichat 6/2, 40127 Bologna, Italy\label{aff23}
\and
Institute for Astronomy, University of Edinburgh, Royal Observatory, Blackford Hill, Edinburgh EH9 3HJ, UK\label{aff24}
\and
Jodrell Bank Centre for Astrophysics, Department of Physics and Astronomy, University of Manchester, Oxford Road, Manchester M13 9PL, UK\label{aff25}
\and
European Space Agency/ESRIN, Largo Galileo Galilei 1, 00044 Frascati, Roma, Italy\label{aff26}
\and
ESAC/ESA, Camino Bajo del Castillo, s/n., Urb. Villafranca del Castillo, 28692 Villanueva de la Ca\~nada, Madrid, Spain\label{aff27}
\and
Universit\'e Claude Bernard Lyon 1, CNRS/IN2P3, IP2I Lyon, UMR 5822, Villeurbanne, F-69100, France\label{aff28}
\and
Institute of Physics, Laboratory of Astrophysics, Ecole Polytechnique F\'ed\'erale de Lausanne (EPFL), Observatoire de Sauverny, 1290 Versoix, Switzerland\label{aff29}
\and
UCB Lyon 1, CNRS/IN2P3, IUF, IP2I Lyon, 4 rue Enrico Fermi, 69622 Villeurbanne, France\label{aff30}
\and
Departamento de F\'isica, Faculdade de Ci\^encias, Universidade de Lisboa, Edif\'icio C8, Campo Grande, PT1749-016 Lisboa, Portugal\label{aff31}
\and
Instituto de Astrof\'isica e Ci\^encias do Espa\c{c}o, Faculdade de Ci\^encias, Universidade de Lisboa, Campo Grande, 1749-016 Lisboa, Portugal\label{aff32}
\and
Department of Astronomy, University of Geneva, ch. d'Ecogia 16, 1290 Versoix, Switzerland\label{aff33}
\and
INAF-Istituto di Astrofisica e Planetologia Spaziali, via del Fosso del Cavaliere, 100, 00100 Roma, Italy\label{aff34}
\and
Universit\'e Paris-Saclay, CNRS, Institut d'astrophysique spatiale, 91405, Orsay, France\label{aff35}
\and
INFN-Padova, Via Marzolo 8, 35131 Padova, Italy\label{aff36}
\and
INAF-Osservatorio Astronomico di Trieste, Via G. B. Tiepolo 11, 34143 Trieste, Italy\label{aff37}
\and
Aix-Marseille Universit\'e, CNRS/IN2P3, CPPM, Marseille, France\label{aff38}
\and
Istituto Nazionale di Fisica Nucleare, Sezione di Bologna, Via Irnerio 46, 40126 Bologna, Italy\label{aff39}
\and
INAF-Osservatorio Astronomico di Padova, Via dell'Osservatorio 5, 35122 Padova, Italy\label{aff40}
\and
Universit\"ats-Sternwarte M\"unchen, Fakult\"at f\"ur Physik, Ludwig-Maximilians-Universit\"at M\"unchen, Scheinerstrasse 1, 81679 M\"unchen, Germany\label{aff41}
\and
Institute of Theoretical Astrophysics, University of Oslo, P.O. Box 1029 Blindern, 0315 Oslo, Norway\label{aff42}
\and
Jet Propulsion Laboratory, California Institute of Technology, 4800 Oak Grove Drive, Pasadena, CA, 91109, USA\label{aff43}
\and
Department of Physics, Lancaster University, Lancaster, LA1 4YB, UK\label{aff44}
\and
Felix Hormuth Engineering, Goethestr. 17, 69181 Leimen, Germany\label{aff45}
\and
Technical University of Denmark, Elektrovej 327, 2800 Kgs. Lyngby, Denmark\label{aff46}
\and
Cosmic Dawn Center (DAWN), Denmark\label{aff47}
\and
Institut d'Astrophysique de Paris, UMR 7095, CNRS, and Sorbonne Universit\'e, 98 bis boulevard Arago, 75014 Paris, France\label{aff48}
\and
Max-Planck-Institut f\"ur Astronomie, K\"onigstuhl 17, 69117 Heidelberg, Germany\label{aff49}
\and
Department of Physics and Helsinki Institute of Physics, Gustaf H\"allstr\"omin katu 2, 00014 University of Helsinki, Finland\label{aff50}
\and
AIM, CEA, CNRS, Universit\'{e} Paris-Saclay, Universit\'{e} de Paris, 91191 Gif-sur-Yvette, France\label{aff51}
\and
Universit\'e de Gen\`eve, D\'epartement de Physique Th\'eorique and Centre for Astroparticle Physics, 24 quai Ernest-Ansermet, CH-1211 Gen\`eve 4, Switzerland\label{aff52}
\and
Department of Physics, P.O. Box 64, 00014 University of Helsinki, Finland\label{aff53}
\and
Helsinki Institute of Physics, Gustaf H{\"a}llstr{\"o}min katu 2, University of Helsinki, Helsinki, Finland\label{aff54}
\and
NOVA optical infrared instrumentation group at ASTRON, Oude Hoogeveensedijk 4, 7991PD, Dwingeloo, The Netherlands\label{aff55}
\and
Dipartimento di Fisica "Aldo Pontremoli", Universit\`a degli Studi di Milano, Via Celoria 16, 20133 Milano, Italy\label{aff56}
\and
INAF-IASF Milano, Via Alfonso Corti 12, 20133 Milano, Italy\label{aff57}
\and
INFN-Sezione di Milano, Via Celoria 16, 20133 Milano, Italy\label{aff58}
\and
Universit\"at Bonn, Argelander-Institut f\"ur Astronomie, Auf dem H\"ugel 71, 53121 Bonn, Germany\label{aff59}
\and
Aix-Marseille Universit\'e, CNRS, CNES, LAM, Marseille, France\label{aff60}
\and
Dipartimento di Fisica e Astronomia "Augusto Righi" - Alma Mater Studiorum Universit\`a di Bologna, via Piero Gobetti 93/2, 40129 Bologna, Italy\label{aff61}
\and
Department of Physics, Centre for Extragalactic Astronomy, Durham University, South Road, DH1 3LE, UK\label{aff62}
\and
Universit\'e C\^{o}te d'Azur, Observatoire de la C\^{o}te d'Azur, CNRS, Laboratoire Lagrange, Bd de l'Observatoire, CS 34229, 06304 Nice cedex 4, France\label{aff63}
\and
Universit\'e Paris Cit\'e, CNRS, Astroparticule et Cosmologie, 75013 Paris, France\label{aff64}
\and
Institut d'Astrophysique de Paris, 98bis Boulevard Arago, 75014, Paris, France\label{aff65}
\and
European Space Agency/ESTEC, Keplerlaan 1, 2201 AZ Noordwijk, The Netherlands\label{aff66}
\and
School of Mathematics, Statistics and Physics, Newcastle University, Herschel Building, Newcastle-upon-Tyne, NE1 7RU, UK\label{aff67}
\and
Department of Physics, Institute for Computational Cosmology, Durham University, South Road, DH1 3LE, UK\label{aff68}
\and
Department of Physics and Astronomy, University of Aarhus, Ny Munkegade 120, DK-8000 Aarhus C, Denmark\label{aff69}
\and
Waterloo Centre for Astrophysics, University of Waterloo, Waterloo, Ontario N2L 3G1, Canada\label{aff70}
\and
Department of Physics and Astronomy, University of Waterloo, Waterloo, Ontario N2L 3G1, Canada\label{aff71}
\and
Perimeter Institute for Theoretical Physics, Waterloo, Ontario N2L 2Y5, Canada\label{aff72}
\and
Space Science Data Center, Italian Space Agency, via del Politecnico snc, 00133 Roma, Italy\label{aff73}
\and
Centre National d'Etudes Spatiales -- Centre spatial de Toulouse, 18 avenue Edouard Belin, 31401 Toulouse Cedex 9, France\label{aff74}
\and
Institute of Space Science, Str. Atomistilor, nr. 409 M\u{a}gurele, Ilfov, 077125, Romania\label{aff75}
\and
Instituto de Astrof\'isica de Canarias, Calle V\'ia L\'actea s/n, 38204, San Crist\'obal de La Laguna, Tenerife, Spain\label{aff76}
\and
Departamento de Astrof\'isica, Universidad de La Laguna, 38206, La Laguna, Tenerife, Spain\label{aff77}
\and
Dipartimento di Fisica e Astronomia "G. Galilei", Universit\`a di Padova, Via Marzolo 8, 35131 Padova, Italy\label{aff78}
\and
Departamento de F\'isica, FCFM, Universidad de Chile, Blanco Encalada 2008, Santiago, Chile\label{aff79}
\and
Institut d'Estudis Espacials de Catalunya (IEEC),  Edifici RDIT, Campus UPC, 08860 Castelldefels, Barcelona, Spain\label{aff80}
\and
Institute of Space Sciences (ICE, CSIC), Campus UAB, Carrer de Can Magrans, s/n, 08193 Barcelona, Spain\label{aff81}
\and
Satlantis, University Science Park, Sede Bld 48940, Leioa-Bilbao, Spain\label{aff82}
\and
Centro de Investigaciones Energ\'eticas, Medioambientales y Tecnol\'ogicas (CIEMAT), Avenida Complutense 40, 28040 Madrid, Spain\label{aff83}
\and
Instituto de Astrof\'isica e Ci\^encias do Espa\c{c}o, Faculdade de Ci\^encias, Universidade de Lisboa, Tapada da Ajuda, 1349-018 Lisboa, Portugal\label{aff84}
\and
Universidad Polit\'ecnica de Cartagena, Departamento de Electr\'onica y Tecnolog\'ia de Computadoras,  Plaza del Hospital 1, 30202 Cartagena, Spain\label{aff85}
\and
Kapteyn Astronomical Institute, University of Groningen, PO Box 800, 9700 AV Groningen, The Netherlands\label{aff86}
\and
INFN-Bologna, Via Irnerio 46, 40126 Bologna, Italy\label{aff87}
\and
Infrared Processing and Analysis Center, California Institute of Technology, Pasadena, CA 91125, USA\label{aff88}
\and
IFPU, Institute for Fundamental Physics of the Universe, via Beirut 2, 34151 Trieste, Italy\label{aff89}
\and
INAF, Istituto di Radioastronomia, Via Piero Gobetti 101, 40129 Bologna, Italy\label{aff90}
\and
Centre de Calcul de l'IN2P3/CNRS, 21 avenue Pierre de Coubertin 69627 Villeurbanne Cedex, France\label{aff91}
\and
INFN-Sezione di Roma, Piazzale Aldo Moro, 2 - c/o Dipartimento di Fisica, Edificio G. Marconi, 00185 Roma, Italy\label{aff92}
\and
Institut f\"ur Theoretische Physik, University of Heidelberg, Philosophenweg 16, 69120 Heidelberg, Germany\label{aff93}
\and
Universit\'e St Joseph; Faculty of Sciences, Beirut, Lebanon\label{aff94}
\and
Junia, EPA department, 41 Bd Vauban, 59800 Lille, France\label{aff95}
\and
SISSA, International School for Advanced Studies, Via Bonomea 265, 34136 Trieste TS, Italy\label{aff96}
\and
INFN, Sezione di Trieste, Via Valerio 2, 34127 Trieste TS, Italy\label{aff97}
\and
ICSC - Centro Nazionale di Ricerca in High Performance Computing, Big Data e Quantum Computing, Via Magnanelli 2, Bologna, Italy\label{aff98}
\and
Instituto de F\'isica Te\'orica UAM-CSIC, Campus de Cantoblanco, 28049 Madrid, Spain\label{aff99}
\and
CERCA/ISO, Department of Physics, Case Western Reserve University, 10900 Euclid Avenue, Cleveland, OH 44106, USA\label{aff100}
\and
Laboratoire Univers et Th\'eorie, Observatoire de Paris, Universit\'e PSL, Universit\'e Paris Cit\'e, CNRS, 92190 Meudon, France\label{aff101}
\and
Dipartimento di Fisica e Scienze della Terra, Universit\`a degli Studi di Ferrara, Via Giuseppe Saragat 1, 44122 Ferrara, Italy\label{aff102}
\and
Istituto Nazionale di Fisica Nucleare, Sezione di Ferrara, Via Giuseppe Saragat 1, 44122 Ferrara, Italy\label{aff103}
\and
Kavli Institute for the Physics and Mathematics of the Universe (WPI), University of Tokyo, Kashiwa, Chiba 277-8583, Japan\label{aff104}
\and
Dipartimento di Fisica - Sezione di Astronomia, Universit\`a di Trieste, Via Tiepolo 11, 34131 Trieste, Italy\label{aff105}
\and
Minnesota Institute for Astrophysics, University of Minnesota, 116 Church St SE, Minneapolis, MN 55455, USA\label{aff106}
\and
Institute Lorentz, Leiden University, Niels Bohrweg 2, 2333 CA Leiden, The Netherlands\label{aff107}
\and
Institute for Astronomy, University of Hawaii, 2680 Woodlawn Drive, Honolulu, HI 96822, USA\label{aff108}
\and
Department of Physics \& Astronomy, University of California Irvine, Irvine CA 92697, USA\label{aff109}
\and
Department of Astronomy \& Physics and Institute for Computational Astrophysics, Saint Mary's University, 923 Robie Street, Halifax, Nova Scotia, B3H 3C3, Canada\label{aff110}
\and
Departamento F\'isica Aplicada, Universidad Polit\'ecnica de Cartagena, Campus Muralla del Mar, 30202 Cartagena, Murcia, Spain\label{aff111}
\and
Department of Physics, Oxford University, Keble Road, Oxford OX1 3RH, UK\label{aff112}
\and
Institute of Cosmology and Gravitation, University of Portsmouth, Portsmouth PO1 3FX, UK\label{aff113}
\and
Department of Computer Science, Aalto University, PO Box 15400, Espoo, FI-00 076, Finland\label{aff114}
\and
Ruhr University Bochum, Faculty of Physics and Astronomy, Astronomical Institute (AIRUB), German Centre for Cosmological Lensing (GCCL), 44780 Bochum, Germany\label{aff115}
\and
Univ. Grenoble Alpes, CNRS, Grenoble INP, LPSC-IN2P3, 53, Avenue des Martyrs, 38000, Grenoble, France\label{aff116}
\and
Department of Physics and Astronomy, Vesilinnantie 5, 20014 University of Turku, Finland\label{aff117}
\and
Serco for European Space Agency (ESA), Camino bajo del Castillo, s/n, Urbanizacion Villafranca del Castillo, Villanueva de la Ca\~nada, 28692 Madrid, Spain\label{aff118}
\and
ARC Centre of Excellence for Dark Matter Particle Physics, Melbourne, Australia\label{aff119}
\and
Centre for Astrophysics \& Supercomputing, Swinburne University of Technology, Victoria 3122, Australia\label{aff120}
\and
School of Physics and Astronomy, Queen Mary University of London, Mile End Road, London E1 4NS, UK\label{aff121}
\and
Department of Physics and Astronomy, University of the Western Cape, Bellville, Cape Town, 7535, South Africa\label{aff122}
\and
ICTP South American Institute for Fundamental Research, Instituto de F\'{\i}sica Te\'orica, Universidade Estadual Paulista, S\~ao Paulo, Brazil\label{aff123}
\and
Oskar Klein Centre for Cosmoparticle Physics, Department of Physics, Stockholm University, Stockholm, SE-106 91, Sweden\label{aff124}
\and
Astrophysics Group, Blackett Laboratory, Imperial College London, London SW7 2AZ, UK\label{aff125}
\and
INAF-Osservatorio Astrofisico di Arcetri, Largo E. Fermi 5, 50125, Firenze, Italy\label{aff126}
\and
Dipartimento di Fisica, Sapienza Universit\`a di Roma, Piazzale Aldo Moro 2, 00185 Roma, Italy\label{aff127}
\and
Centro de Astrof\'{\i}sica da Universidade do Porto, Rua das Estrelas, 4150-762 Porto, Portugal\label{aff128}
\and
Dipartimento di Fisica, Universit\`a di Roma Tor Vergata, Via della Ricerca Scientifica 1, Roma, Italy\label{aff129}
\and
INFN, Sezione di Roma 2, Via della Ricerca Scientifica 1, Roma, Italy\label{aff130}
\and
Institute of Astronomy, University of Cambridge, Madingley Road, Cambridge CB3 0HA, UK\label{aff131}
\and
Department of Astrophysics, University of Zurich, Winterthurerstrasse 190, 8057 Zurich, Switzerland\label{aff132}
\and
Dipartimento di Fisica, Universit\`a degli studi di Genova, and INFN-Sezione di Genova, via Dodecaneso 33, 16146, Genova, Italy\label{aff133}
\and
Theoretical astrophysics, Department of Physics and Astronomy, Uppsala University, Box 515, 751 20 Uppsala, Sweden\label{aff134}
\and
Department of Physics and Astronomy, University College London, Gower Street, London WC1E 6BT, UK\label{aff135}
\and
Department of Astrophysical Sciences, Peyton Hall, Princeton University, Princeton, NJ 08544, USA\label{aff136}
\and
Niels Bohr Institute, University of Copenhagen, Jagtvej 128, 2200 Copenhagen, Denmark\label{aff137}
\and
Cosmic Dawn Center (DAWN)\label{aff138}
\and
Center for Cosmology and Particle Physics, Department of Physics, New York University, New York, NY 10003, USA\label{aff139}
\and
Center for Computational Astrophysics, Flatiron Institute, 162 5th Avenue, 10010, New York, NY, USA\label{aff140}}

\date{}

\titlerunning{\Euclid preparation: multi-tracer method with  the  spectroscopic and photometric samples 
}
\authorrunning{F. Dournac et al.}
 
%
%
\abstract{
    Future data provided by the \Euclid mission will allow us to better understand the cosmic history of the Universe. A metric of its 
    performance is the  {figure-of-merit} (FoM) of dark energy, usually estimated with Fisher forecasts. The expected FoM has previously been  estimated taking into account the two main probes of \Euclid, namely the three-dimensional clustering of the 
    spectroscopic galaxy sample, and the so-called {3$\times$2}\,pt signal from 
    the photometric sample (i.e., the weak lensing signal, the galaxy clustering, 
    and their cross-correlation). So far, these two probes have been 
    treated as independent.
    In this paper, we introduce a new observable given by the ratio of the (angular) two-point correlation function of galaxies from 
    the two surveys.  For identical (normalised) selection functions, this observable is unaffected by sampling noise, and its variance is solely controlled by Poisson noise. We present forecasts for \Euclid where this multi-tracer method is applied and is particularly relevant {because} the two surveys will  cover the same area of the sky.
    This method allows for the exploitation of the combination {of} the spectroscopic and photometric samples.
     When the correlation between this new observable and the other probes is not taken into account, a significant gain is obtained in the FoM, as well as in the constraints on other cosmological parameters. 
    The benefit is more pronounced for {a commonly} investigated modified gravity model, namely the $\gamma$ parametrisation of the growth factor.
    However, the correlation between the different probes is found to be significant {and hence the actual gain is uncertain}. 
    We present various strategies for circumventing this issue and still extract useful information from the new observable. 
}
%
%
%
%
   
\maketitle
%
%
%
%
\section{Introduction}
    In recent decades, a large number of observations and studies have been converging on the fact that our Universe is going through a phase of accelerated expansion,  visible on cosmological scales. In order to better understand the origin of this cosmic acceleration and the physics of gravity on large scales, wide galaxy surveys such as \Euclid\footnote{\url{https://www.euclid-ec.org}}~\citep{laureijs2011euclid} rely essentially on two main probes: galaxy clustering, denoted here by $\GCsp$ ($\GCph$) for analyses with spectroscopic (photometric) redshifts; and weak gravitational lensing (WL), also known as cosmic shear. The $\GCsp$ and $\GCph$ probes aim at reconstructing the fluctuations of the underlying dark matter density using coordinates and redshifts from the angular and radial positions of galaxies. Different measurements can be done to extract information from this underlying distribution, such as measurements of the baryon acoustic oscillations~\citep[BAOs;][]{2005ApJ...633..560E,2015PhRvD..92l3516A}, or measurements of the redshift-space distortion effects \citep[RSD;][]{2009MNRAS.393..297P}. 
    Complementary to the clustering probes, galaxy surveys enable $\WL$ analyses. They characterise the matter present along the line of sight, which slightly alters the images of galaxies as a function of the gravitational potentials  traversed by photons~\citep[see, e.g.,][for a detailed review]{Kilbinger_review}. WL analyses  not only extract information about the matter content of the Universe, but also about the growth of structure and the physics of gravitational interaction. Stage-IV galaxy surveys, such as \Euclid, will provide a large amount of data that will enable very precise $\GCsp$, $\GCph$, and $\WL$ analyses~\citep[see, e.g.,][from hereafter EP:VII]{IST:paper1}.
    
    As shown in~\citetalias{IST:paper1}, the combination of all main probes ($\GCsp$, $\GCph$, and WL) will lead to the most stringent constraints from future \Euclid data. The combination of different probes, sensitive to different aspects of how gravity acts in the cosmos, breaks several degeneracies present between the different cosmological parameters and achieves better constraints. 

    However, such cosmological probes are in general not independent. It was shown in~\citetalias{IST:paper1} that the cross-correlations between $\GCph$ and WL were another important key contributor in the joint analysis of all \Euclid probes. More precisely, the figure-of-merit \citep[FoM,][]{2006astro.ph..9591A} of the dark energy equation-of-state parameters\,\footnote{We use the {alternative} definition {from \cite{2008PhRvD..77l3525W} }FoM=$\sqrt{\text{det}\left(\tilde{F}_{w_0w_a}\right)}$, with $\tilde{F}_{w_0w_a}$ being the marginalised Fisher submatrix for the dark energy equation-of-state parameters.} improves roughly by a factor of 4 when these cross-correlations are included in the analysis. It was also shown in~\citet{Tutusaus_2020} that a joint analysis of \Euclid photometric probes accounting for their cross-correlations can significantly improve our knowledge on nuisance parameters, such as the intrinsic alignment of galaxies or the galaxy bias. Cross-correlations between $\GCph$ and WL have been studied for real observations~\citep[see, e.g.,][]{DESY3} and the future \Euclid data~\citep{Tutusaus_2020}. Similarly, there have been several analysis combining spectroscopic and photometric data\,\citep[see, e.g.,][]{2021A&A...646A.140H,2023arXiv230400705S}. However, the full treatment of all cross-correlations between the spectroscopic probe, $\GCsp$, and the photometric probes, taking into account their covariances and the radial information, has been less considered. One of the main reasons is that spectroscopic analyses are usually performed in three-dimensions, while photometric analyses are done in two-dimensions. This difference makes it non-trivial to properly combine the spectroscopic and photometric probes while accounting for their cross-correlations~\citep[although several attempts are available in the literature, see, e.g.,][]{sphericalFB,sphericalFB2,PeterDida}. In~\citetalias{IST:paper1}, the authors neglected any correlation between $\GCsp$ and the photometric probes. In the case of \Euclid, another motivation for this choice is that the spectroscopic measurements will  be available only at high redshift ($z>0.9$), therefore reducing their overlap in volume with the photometric probes. In order to be conservative, a pessimistic scenario was further considered in~\citetalias{IST:paper1} where all objects above $z=0.9$ were removed for $\GCph$ (and their cross-correlations with WL), with the goal of removing any remaining correlation. 
    
    In the present analysis we go beyond the results presented in~\citetalias{IST:paper1} by focusing on extracting additional information from the combination of spectroscopic and photometric probes. Without accounting for all the cross-correlations between spectroscopic and photometric observables (which would require a joint modelling of three- and two-dimensional probes), we extract the additional information from the fact that $\GCsp$ and $\GCph$ will probe (at least partially) the same volume of the Universe. 
    
    To do so, we introduce the ratio of angular correlation functions (or harmonic space power spectra) between the spectroscopic and photometric tracers as an additional observable { according to  \cite{1988A&A...206L..11A} who first suggested this type of statistic to get rid of {sampling} variance. This approach  made it possible to implement a multi-tracer approach \citep{PhysRevLett.102.021302, 2009JCAP...10..007M}. Indeed, } given the large number density of \Euclid objects, if tracers are accurately selected we can get rid of the cosmic variance and obtain very precise measurements of these ratios. In practice, this implies that once the bias of one tracer is known, there is effectively almost no uncertainty on the bias of the other tracer. Given this reduction of the number of degrees of freedom, we can significantly improve the \Euclid forecasts. In the end, this leads to an improvement of the FoM ranging from 5\,\% up to 60\,\%, 
    compared to the baseline analysis presented in~\citetalias{IST:paper1}. The specific improvement depends on the settings {of the surveys} and the cosmological model considered.
\Euclid is an ESA M-class space mission whose main goal is to measure the geometry of the Universe and the growth of structures {out} to redshift $z\sim 2$, i.e., a look-back time of 10Gyr and beyond. This space telescope, launched on  1st July 2023, carries a near-infrared spectrometer and photometer instrument~\citep[\Euclid-NISP]{2022SPIE12180E..1KM} and a visible imager~\citep[\Euclid-VIS]{VIS_paper}. These two detectors will perform a galaxy survey over an area of about $15\,000\,\deg^2$ of the extragalactic sky. \Euclid-NISP will be able to measure  30 {to} 50 million spectroscopic redshifts between 0.9 and 1.8~\citep{Pozzetti2016}, which can be used for GC measurements, while \Euclid-VIS will measure about 1.5 billion photometric galaxy shapes, enabling weak lensing (WL) observations \citep[see][for more details]{laureijs2011euclid}. The huge volume of data provided by \Euclid will give  new insight{s} into the late Universe, especially on the growth and evolution of large-scale cosmic structures and on the expansion history of the Universe, and more generally shed some light on the nature of dark energy \citep[see][]{2018LRR....21....2A}.

    The paper is organised as follows.  We first describe the \Euclid survey and how we forecast its constraining power for the main cosmological probes in Sect.\,\ref{sec:2}. We then present the cosmological models considered in Sect.\,\ref{sec:3}. In Sect.\,\ref{sec:4} we introduce our new observable making use of the multi-tracer approach, and clarify in Sect.\,\ref{sec:5} its implementation in our forecasts. The main results of the analysis are presented in Sect.\,\ref{sec:6}, and we conclude in Sect.\,\ref{sec:7}.
    
\section{The main \Euclid cosmological probes}\label{sec:2}
In this section we describe how we forecast the constraining power of \Euclid for its main cosmological probes. We follow closely the recipes presented in~\citetalias{IST:paper1} and, although we provide a self-contained description in this work, we refer the interested reader to~\citetalias{IST:paper1} for all the technical details.

\subsection{Spectroscopic probe}\label{sec:2.2}

Let us first consider the spectroscopic probe of \Euclid. The main observable for this probe is the observed galaxy power spectrum, which needs a reference cosmology. Following \citetalias{IST:paper1} we model it as
\begin{multline}
P_\text{obs}(k_{\rm ref}, \mu_{\rm ref} ;z) = 
\frac{1}{q_\perp^2(z)\, q_\parallel(z)} 
\left\{\frac{\left[b_{\rm sp}\sigma_8(z)+f\sigma_8(z)\mu^2\right]^2}{1+ \Big[ f(z) k \mu \sigma_{\rm p}(z) \Big]^2 } \right\} 
\\
\times \frac{P_\text{dw}(k,\mu;z)}{\sigma_8^2(z)}  
F_z(k,\mu;z) 
+ P_\text{s}(z) \,, 
\label{eq:GC:pk-ext}
\end{multline}
where $\sigma_8$ is the r.m.s. of linear matter fluctuations on scales of 8\,$h$$^{-1}$Mpc, $b_{\rm sp}$ the galaxy bias parameter, $f$ the growth rate, $ \sigma_{\rm p}$ is related to the pairwise peculiar velocity and treated as a free nuisance parameter, $\mu$ the cosine of the angle between the wave vector, $\vec{k}$, and the line-of-sight direction, and all $k:=k(k_{\rm ref},\mu_{\rm ref})$ and $\mu:=\mu(\mu_{\rm ref})$ with

\begin{align}
    k(k_{\rm ref},\mu_{\rm ref})&=\frac{k_{\rm ref
}}{q_{\perp}}\left[1+\mu_{\rm ref}^2\left(\frac
{q_{\perp}^2}{q_{\parallel}^2}-1\right)\right]^{1/2}\,,\\
\mu(\mu_{\rm ref}) &= \mu_{\rm ref}\frac{q_{\perp}}{q_{\parallel}}\left[1+\mu_{\rm ref}^2\left(\frac
{q_{\perp}^2}{q_{\parallel}^2}-1\right)\right]^{1/2}\,.
\end{align}

RSDs are accounted for through the numerator inside the curly bracket in Eq.~(\ref{eq:GC:pk-ext}), and the 'finger-of-God' effect through the denominator. 
The term $P_{\rm dw}$ is the 'de-wiggled' power spectrum, which accounts for the smearing of the BAOs:
\begin{equation}
P_\text{dw}(k,\mu;z) = P_{\delta\delta}^{\rm lin}(k;z)\,\text{e}^{-g_\mu k^2} + P_\text{nw}(k;z)\left(1-\text{e}^{-g_\mu k^2}\right) \,,
\label{eq:pk_dw}
\end{equation}
where $P_{\rm nw}$ stands for a no-wiggle power spectrum with the same broad band {shape }as the linear power spectrum, $P_{\delta\delta}^{\rm lin}$, but without the BAO wiggles. We also account for nonlinearities through a nonlinear damping factor
\begin{equation}
g_\mu(k,\mu,z) = \sigma_{\rm v}^2(z) \left\{ 1 - \mu^2 + \mu^2 [1 + f(z)]^2  \right\}\,,
\end{equation}
with
\begin{equation}
\sigma_{\rm v}^2 (z) = \frac{1}{6\pi^2} \int_0^{+\infty} \text{d} k \; P_{\delta \delta }^{\rm lin} (k,z) \,.
\end{equation}

 We additionally introduce
\begin{equation}
    F_z(k, \mu;z) = \text{e}^{-k^2\mu^2\sigma_{r}^2(z)}\,,
\end{equation}
where $\sigma_{r}(z) = (1+z)\sigma_{z} c/H(z)$ accounts for redshift uncertainties and is modelled as  in~\citetalias{IST:paper1}. We set $\sigma_z=10^{-3}$. Moreover, we consider a residual shot-noise $P_{\rm s}$ as a {constant}
 nuisance parameter {in each redshift bin}. Finally, the quantities
\begin{align}
q_{\perp}(z) &= \frac{r(z)}{r_{\rm ref}(z)},\\
q_{\parallel}(z) &= \frac{H_\text{ref}(z)}{H(z)}\,,
\end{align}
account for the Alcock-Paczynski effect \citep{1979Natur.281..358A}, where $r(z)$ stands for the comoving angular distance and $H(z)$ is the Hubble expansion rate. The quantities indexed by 'ref' refer to the quantities in the reference cosmology required for measurements of the power spectrum in the spectroscopic survey.

\subsection{Photometric probes}

With respect to the photometric survey, we consider the harmonic-space angular power spectra $C_{ij}^{XY}(\ell)$, with $i$ and $j$ representing two tomographic bins, and $X$ and $Y$ being either GC$_{\rm ph}$ or WL. Under the {extended} Limber approximation \citep{2008PhRvD..78l3506L}, these are given by
\begin{equation}
 C^{XY}_{ij}(\ell) = c\int_{z_{\rm min}}^{z_{\rm max}}\text{d} z\,{\frac{W_i^X(z)W_j^Y(z)}{H(z)r^2(z)}P_{\delta\delta}(k_\ell,z)}\,, \label{eq:ISTrecipe}
\end{equation}
where $k_\ell=(\ell+1/2)/r(z)$, and the nonlinear matter power spectrum is represented by $P_{\delta\delta}$.

The remaining ingredients in Eq.\,(\ref{eq:ISTrecipe}) are the kernels for GC$_{\rm ph}$ or WL,
\begin{align}
 W_i^{\rm GC_{\rm ph}}(z) =&\; b_i^{\rm {ph}}(z)n_i(z)\frac{H(z)}{c}\,, \\  
 W_i^{\rm WL}(z) =&\; \frac{3}{2}\Omegam \frac{H_0^2}{c^2}(1+z)\,r(z)
\int_z^{z_{\rm max}}{\text{d} z'{n_i(z')}%
\frac{r(z,z')}{r(z')}}\nonumber\\
  &+W^{\rm IA}_i(z)\,, 
\end{align}
with {$n_i(z')$} 
being the normalised number density {distribution} of galaxies in {tomographic} bin $i$, $b_i^{\rm {ph}}(z)$  the {linear} galaxy bias in {the same bin}, and $r(z,z')$ the comoving angular {diameter} distance of a source at redshift $z'$ seen from an observer at redshift $z$. We also consider the intrinsic alignment (IA) of galaxies with the extended nonlinear alignment (eNLA) model{, as in \citetalias{IST:paper1}. This corresponds to the kernel
\begin{equation}\label{eq:IA}
 W^{\rm IA}_i(k,z)=-\frac{\mathcal{A}_{\rm IA}\,\mathcal{C}_{\rm IA}\,\Omega_{\rm m,0}\,\mathcal{F}_{\rm IA}(z)}{{D(z)}}
\frac{H(z)}{c}\,,
\end{equation}
with $D$ the growth factor for linear perturbations:
\begin{equation}
\delta(z)=  D(z)\, \delta_0\,,
\end{equation}
normalised to 1 at $z=0$, 
and
\begin{equation}
 \mathcal{F}_{\rm IA}(z)=(1+z)^{\eta_{\rm IA}}\left[\frac{\langle L\rangle(z)}{L_\ast(z)}\right]^{\beta_{\rm IA}}\,.
\end{equation}
The functions $\langle L\rangle(z)$  and $L_\ast(z)$  are the redshift-dependent mean and the characteristic luminosity of source galaxies as computed from the luminosity function. For the parameters of the IA model, $\eta_{\rm IA}$, $\beta_{\rm IA}$, $\cal{C}_{\rm IA}$, and $\cal{A}_{\rm IA}$, {we consider the}  
fiducial values {presented}  
in~\citetalias{IST:paper1}.
Further details on the eNLA model and the luminosity dependence assumed can be found there.

\subsection{Forecast code}
In order to compute the observables described in the previous sections and forecast their cosmological power, we consider here the \texttt{TotallySAF}\,\footnote{\url{https://github.com/syahiacherif/TotallySAF_Alpha}} code, which relies on \texttt{CAMB}\,\citep{2000ApJ...538..473L} to solve the Boltzmann equations. \texttt{TotallySAF} has been used previously for forecasting the constraining power of the main cosmological probes of \Euclid using the Fisher formalism~\citepalias{IST:paper1}. The same code allows us to forecast both the spectroscopic (GC$_{\rm sp}$) and photometric probes (WL and GC$_{\rm ph}$), as well as their cross-correlations. An important feature of this code is the possibility to specify the number of points in the $n$-point stencil derivatives and therefore achieve a high level of accuracy, avoiding numerical instabilities in the Fisher forecast \citep{RefSafirpaper}.

\section{Cosmological models}\label{sec:3}

In the present study, the cosmological models investigated are the ones described in \citetalias{IST:paper1}. These models are spatially flat universes filled with cold dark matter and dynamical dark energy. {We also considered non-flat models and a modified gravity model}. The dynamics of dark energy is described by a time-varying equation-of-state {parameter} following the popular CPL parameterisation 
\citep{Chevallier_Polarski_2001,2005PhRvD..72d3529L}:
\begin{equation}
w(z)=w_0+w_a\frac{z}{1+z}\,.
\label{eq:cpl}
\end{equation}

Five  
other  cosmological parameters enter in  the model:
the dimensionless Hubble constant $h$ (defined by $H_0 = 100 h$\,km\,s$^{-1}$\,Mpc$^{-1}$); the total matter density at {the} present time $\Omega_{\rm m}$;
the dark energy density at {the} present time $\Omega_{\rm DE}$; the current baryonic matter density $\Omega_{\rm b}$; the spectral index $n_{\rm s}$ of the primordial spectrum of scalar perturbations; and the current amplitude of matter fluctuations as expressed by $\sigma_{8}$ (the r.m.s. of linear matter fluctuations in spheres of $8 h^{-1}$\,Mpc radius).  For the description of linear matter perturbations in \texttt{CAMB} we take into account here the  parameterised-post-Friedmann (PPF) framework of \citet{Hu:2007pj}, which enables the equation-of-state to cross $w(z)= -1$ without developing instabilities in the perturbation sector. 

Extensions of $\Lambda$CDM theories may alter the background {as well as} the perturbation sector. A simple way to investigate this possibility is through 
the growth index $\gamma$, defined as 
\begin{equation}
\gamma(z) = \frac{\ln f(z)}{\ln\Omega_{\rm m}(z)}\,, \quad\text{with}\quad \Omega_{\rm m}(z)\equiv\dfrac{\Omegam(1+z)^3H_0^2}{H^2(z)}\,,
\label{eq:fgrowth}
\end{equation}
{and $f(z)$} the growth  rate, 
\begin{equation}
f(z)\equiv-\dfrac{\text{d}\ln D(z)}{\text{d}\ln(1+z)}\,.
\end{equation}

In standard gravity models, the growth factor is well approximated with a constant $\gamma \approx 0.55$, having a weak dependence on $\Lambda$ \citep{1991MNRAS.251..128L}. Introducing $\gamma$ as a constant free parameter is a simple way to describe possible departures from the standard model \citep{2003PhRvL..90i1301L}.

The fiducial case is the standard concordance model, that is, 
a spatially-flat Universe filled mostly with cold dark matter and a cosmological constant. Our cosmological models are described by the following vector of parameters with their fiducial values coming from \citetalias{IST:paper1}:
\begin{align}\label{eq.paramsfid}
\vec{\lambda}&=\{\Omegam,\,\Omegab,\,w_0,\,w_a,\,h,\,n_{\rm s},\,\sigma_{8},\,\Omega_{{\rm DE}},\,\gamma\}\nonumber\\
&=\{0.32,\,0.05,\,-1.0,\,0.0,\,0.67,\,0.96,\,0.816,\,0.68,\,0.55\}\,.
\end{align}

There are also {several} 
nuisance parameters. 
For the photometric sample, these include the {three parameters for the} intrinsic alignment{s, $A_{\rm IA}, \eta_{\rm IA}, \beta_{\rm IA}$, and the linear galaxy bias in each tomographic bin.} 
For the spectroscopic sample, {we consider the linear galaxy bias parameter and the residual shot noise, $P_{\rm s}$, in each redshift bin.}

The sum of neutrino masses is also fixed to $0.06\,$eV. In the presence of massive neutrinos, the redshift and scale dependence of the linear growth factor differs from the zero-mass case. This effect is taken into account in standard Boltzmann solvers such as \texttt{CAMB} \citep{2000ApJ...538..473L}or \texttt{CLASS} \citep{2011arXiv1104.2932L}. However, 
given the small neutrino masses considered in this work, we neglect neutrino effects on the growth factor, following the same   approach as in \citetalias{IST:paper1}, for simplicity.

\section{The ratio of  correlation functions as  additional information}\label{sec:4}
The ratio of the correlation functions of two different galaxy samples, also known as the ratio of cross-correlations, is a powerful method to measure the ratio of the galaxy bias of their respective populations \citep{1988A&A...206L..11A}.
Let us consider measuring two galaxy populations, tracers  of the matter density field, $\delta_1$ and $\delta_2$. {Let us further assume} that both  tracers are Poisson realizations of the underlying  density fields  $\delta_i = b_i\delta$,  {with} $b_1$, $b_2$ being the large-scale biases of the two tracers. When the two galaxy populations follow the same selection function over the same volume, this ratio is insensitive to the sampling variance and the Poisson noise is the only source of variance. {If the selection functions are different but known, an appropriate weighting scheme will achieve the same result.} This is the essence of the multi-tracer approach. It can also be applied to RSD {and non-gaussianity} measurements \citep{PhysRevLett.102.021302, 2009JCAP...10..007M}. This approach can be applied equivalently to the angular correlation functions or the harmonic power spectra of {the} two galaxy samples \citep[see for instance][]{2021MNRAS.502.2952T,2022JCAP...08..073A}.   
In the following, we will consider these two populations to be the spectroscopic and photometric populations of the galaxies observed by \Euclid. {However, it is important to have the same selection function in both data sets in order to benefit from the insensitivity to sample variance. Because of this, we will choose the spectroscopic sample by selecting those galaxies in each one of the photometric tomographic bins for which spectroscopic information is available. 
}  This is aimed at ensuring that the selection function for both data sets in the new observable will be the same, that is, the photometric selection function. A weighting scheme could be necessary to properly achieve this goal, which is feasible as long as the selection functions of the samples are known. {To ensure that the same galaxies are not used twice, we will assume that the galaxies in the spectroscopic sample have been removed from the photometric sample. }

We first denote as $a_{\ell m,\rm sp}$ ($a_{\ell m,\rm ph}$)  the coefficients of the spherical-harmonic decomposition of the spectroscopic (photometric) galaxy distribution and the corresponding angular power spectrum $C_{\ell m,\rm sp}$ ($C_{\ell m,\rm ph}$). In the absence of any Poisson noise, and assuming a linear galaxy bias relation, we have
\begin{equation}
\dfrac{|a_{\ell m,\rm sp}|^2}{|a_{\ell m,\rm ph}|^2} = \dfrac{C_{\ell,\rm sp}}{C_{\ell,\rm ph}} =\left(\dfrac{b_{\rm sp}}{b_{\rm ph}}\right)^2\,,
\end{equation}
where $b_{\rm sp}$ ($b_{\rm ph}$) stands for the {linear} galaxy bias of the spectroscopic (photometric) population.

We consider a spectroscopic sample over a finite volume {(given by the photometric selection)}, assumed to be a  Poisson realisation of a field with a density $n_{\rm sp}$, which represents the galaxy surface  density of the spectroscopic sample (in inverse steradians), as well as an unbiased estimator $\widehat{|a_{\ell m,\rm sp}|}{}^2$ of $|a_{\ell m,\rm sp}|^2$. Over the same volume, with an identical selection function, we also consider an unbiased estimation of $|a_{\ell m,\rm ph}|^2$.
We can express the ratio $o_{\ell }$ of the angular power spectrum of the two populations as an observable quantity, an estimator of which is
\begin{equation}
\widehat{o_{\ell }} = \frac{1}{2l+1} \sum_m\dfrac{\widehat{|a_{\ell m,\rm sp}|}{}^2}{\widehat{|a_{\ell m,\rm ph}|^2}}\,,
\label{eq:observable}
\end{equation}
 For the sake of simplicity, we assume that the Poisson noise for the photometric sample can be neglected, as $n_{\rm ph}$ is expected to be much larger than $n_{\rm sp}$ in \Euclid. {This means that the shot noise is assumed to be low compared to the spectrum itself. Appropriate processing may be required for data applications (see for example~\citet{2017MNRAS.471L..57T}). 
}
Its average over Poisson realisations  is then
\begin{equation}
\overline{o_{\ell}}\equiv\left<\widehat{o_{\ell}}\right> = \left<\dfrac{\widehat{C_{\ell,\rm sp}}}{C_{\ell,\rm ph}}\right> = \dfrac{\left<\widehat{C_{\ell,\rm sp}}\right>}{C_{\ell,\rm ph}} = \left(\dfrac{b_{\rm sp}}{b_{\rm ph}}\right)^2\,.
\label{eq:mean value}
\end{equation}
Since this quantity is a constant independent of the sample, the average (over samples) is also the same (notice that in this expression $C_l$ is the realization on the specific survey  and differs from  the ensemble average). The  variance $\sigma_o^2(\ell)$ of $\widehat{o_{\ell}}$  can be inferred {(see \hyperref[Append_label]{Appendix})} :

\begin{equation}\label{eq:variance o_ell}
\sigma_o^2(\ell)=\frac{1}{2\ell+1}\left(
\frac{4 C_{\ell,\rm sp}}{f_{\rm sky}\,C_{\ell,\rm ph}^2\,n_{\rm sp}}
+\frac{2}{f_{\rm sky}\,C_{\ell,\rm ph}^2\,n_{\rm sp}^{2}}\right)\,. 
\end{equation}

{The parameter} $f_{\rm sky}$ represents the fraction of sky observed by \Euclid. 
We can thus build an estimator $\hat{O}$ of $ \displaystyle \left(\frac{b_{\rm sp}}{b_{\rm ph}}\right)^2$ by taking the optimal (inverse-variance weighted) average over all $\ell$ and~$m$ :
\begin{equation}
\displaystyle
 \hat{O} = 
 \dfrac{\sum\limits_{\ell=\ell_{\rm min}}^{\ell_{\rm max}}  \dfrac{1}{\sigma_o^2(\ell)} \widehat{o_{\ell }}}{\sum\limits_{\ell=\ell_{\rm min}}^{\ell_{\rm max}}  \dfrac{1}{\sigma_o^2(\ell)}}
 \,. 
\label{first_o}
\end{equation}
The values $\ell_{\min}$ and $\ell_{\max}$ depend on the scenario and are specified in Sect.\,\ref{section:results} in our case. 
The variance of this new observable $\hat{O}$ can then be written as :

\begin{equation}                        
\sigma_{O}^{2}=\left(\sum_{\ell=\ell_{\min }}^{\ell_{\max }} \frac{(2 \ell+1) C_{\ell, \mathrm{ph}}^2}{1+2 n_{s p} C_{\ell, \mathrm{sp}}}\right)^{-1} \frac{2}{f_{\mathrm{sky}} n_{\mathrm{sp}}^2}   \,. 
\label{variance_o3_final}
\end{equation}

{It is important to note that we require the same selection function for both the spectroscopic and photometric data sets to cancel out the dependence on sample variance in Eq.\,(\ref{eq:mean value}) and {obtain} a direct link between this new observable and the ratio of the linear galaxy biases. That being said, we still consider the standard modelling for GC$_{\rm sp}$ presented in \citetalias{IST:paper1} and summarized in Sect.\,\ref{sec:2.2}. For example, we include the impact of RSDs and the finger-of-God effect when modelling the spectroscopic probe. In practice, we consider two different spectroscopic selections: the standard one, with narrow redshift bins; and a new one, derived from the photometric selection, with broad bins. We use the former to derive constraints from GC$_{\rm sp}$, like in \citetalias{IST:paper1}, while we only consider the latter to constrain the ratio of the linear galaxy biases present in Eq.\,(\ref{eq:mean value}).} {We rely on two {basic} approximations in {using} Eq.\,(\ref{eq:mean value}). The first one is to assume that only the density term is important for galaxy number counts when we consider the broad photometric selection function, as was considered in \citetalias{IST:paper1}, for simplicity. Although other terms, like RSD might have a non-negligible contribution\,\citep[see, e.g,][]{Tanidis}, the change in constraining power when including these effects is very small. Therefore, the same justification to neglect these terms for the photometric data set hold for neglecting them for the harmonic power spectra from the spectroscopic data set (with the broad selection function). The second approximation {is that we assume that the linear galaxy bias for the spectroscopic sample used in GC$_{\rm sp}$ (and therefore considering a top-hat selection function) is the same as the linear galaxy bias for the spectroscopic sample used in the new observable XC2, which considers the photometric selection function for the spectroscopic sample, too}. In reality, the linear galaxy bias $b_{\rm sp}$ present in Eq.\,(\ref{eq:mean value}) might depend on the selection function and be slightly different compared to the linear galaxy bias that enters Eq.\,(\ref{eq:GC:pk-ext}) to model the observed power spectra. Given that both the narrow and broad selection function for the spectroscopic data are centred at the same effective redshift, we assume these two parameters to be the same. {We have checked that the difference is smaller than 2\% in our cases. In practice, a large difference could however be taken into account if necessary by evaluating from the data the ratio of biases for both samples.}}

\section{Introducing the new observable in Fisher forecasts}\label{sec:5}

\subsection{Computing the additional Fisher matrix}

We recall briefly the Fisher matrix formalism for a given likelihood $\mathcal{L}$ and model parameters vector $\vec{\lambda}=\{\lambda_i\}$ (typically cosmological parameters) with fiducial values $\lambda_{i,{\rm fid}}$. The $F_{ij}$ element (where the indices $i$ and $j$ run over model parameters) of the Fisher matrix $F$ is defined as
\begin{equation}
F_{i j}\equiv \ave{ \left . -\dfrac{\partial^2 \ln (\mathcal{L})}{\partial \lambda_{i} \partial\lambda_{j}} \right|_{\lambda=\lambda_{\rm fid}}  } 
,
\label{fisher}
\end{equation}
\citep{2017arXiv170501064L} where brackets denote an ensemble average over all possible realisations of the observables considered, given our fiducial model. Assuming that the vector of observables follows a Gaussian distribution, with mean $\mu$ and covariance \tens{C} (which both can depend on the model parameters), the Fisher matrix can be written analytically as
\begin{equation}\label{eq_general_fisher}
    F_{ij}=\frac{1}{2}\,\text{tr}\left[\frac{\partial \tens{C}}{\partial \lambda_i}\tens{C}^{-1}\frac{\partial \tens{C}}{\partial \lambda_j}\tens{C}^{-1}\right]+\sum_{mn}\frac{\partial \mu_m}{\partial \lambda_i}(\tens{C}^{-1})_{mn}\frac{\partial \mu_n}{\partial \lambda_j}\,.
\end{equation}

We first consider the combination of two probes $A$ and $B$; as an example, one may consider the spectroscopic probe ($A=\GCsp$) and the combination of photometric galaxy clustering, weak lensing, and their cross-correlation ($B=\GCph+\WL+\XC$, often referred to as 3$\times$2\,pt in the literature). Defining their respective likelihoods as $\mathcal{L}_A=\mathcal{L}(\GCsp)$ and $\mathcal{L}_B=\mathcal{L}(\GCph+\WL+\XC)$, and assuming that $A$ and $B$ are not correlated, then the combined likelihood of both probes will simply be given by the product of the individual likelihoods.

The main idea of the present work is to go beyond this simple probe combination by exploiting the fact that the two probes $A$ and $B$ share the same volume. To do so we introduce the observable $O$, the ratio of correlation functions defined in Sect.~\ref{sec:4} {through} Eq.~(\ref{first_o}). Assuming that this new observable (with associated likelihood $\mathcal{L}_O$) is also independent of $A$ and $B$ (the role of correlations is discussed in Sect.\,\ref{correlation_section}), the total likelihood is
\begin{equation}
\mathcal{L}(\GCsp+\GCph+\WL+\XC+O)=\mathcal{L}_A\,\mathcal{L}_B\,\mathcal{L}_O\,.
\end{equation}

This new likelihood $\mathcal{L}_O$ is associated with a 'new' data vector of dimension 
equal to the number of overlapping redshift bins 
between the spectroscopic and photometric probes. More details  are provided in Sect.\,\ref{sec61} on the binning of the two probes to ensure the same selection function in the case of \Euclid. In each redshift bin, the mean value $\mu_{O}=(b_{\text{sp}}/b_{\text{ph}})^2$ and variance $\sigma_{O}^2$ of those new observables are obtained respectively from Eqs.~(\ref{eq:mean value}) and~(\ref{variance_o3_final}) in Sect.~\ref{sec:4}. The new contribution $F^O_{ij}$ to the total Fisher matrix can then be computed for each redshift, thanks to Eq.~(\ref{eq_general_fisher}),

\begin{equation}
F^O_{ij}=\dfrac{1}{\sigma_{O}^{2}}\left(\dfrac{\partial \mu_{O}}{\partial \lambda_i} \dfrac{\partial \mu_{O}}{\partial \lambda_j}\right)\,,
\label{fisher_o}
\end{equation}
where we assume that the observable $O$ follows a Gaussian distribution, and that we ignore the  dependence of its variance {on}  the cosmological and nuisance parameters.

The mean value $\mu_{O}$ of the observable $O$ 
does not depend on the cosmological parameters, but on  
$b_{\text{sp}}$ and $b_{\text{ph}}$. Therefore, the only nonzero partial derivatives are
\begin{equation}
    \frac{\partial \mu_{O}}{\partial b_{\text{sp}}}=\frac{2\,b_{\text{sp}}}{b_{\text{ph}}^{2}} \quad\text{and}\quad \frac{\partial \mu_{O}}{\partial b_{\text{ph}}}=-\frac{2\,b_{\text{sp}}^{2}}{b_{\text{ph}}^{3}}\,.
\end{equation}
As a consequence, the only nonzero elements of $F^{O}$ reduce to a 2$\times$2 matrix for each redshift bin, given by
    \begin{equation}
F^{O}=\dfrac{1}{\sigma_{O}^{2}}\left(\begin{array}{cc}
\left(\dfrac{\partial \mu_{O}}{\partial b_{\text{sp}}}\right)^2 & \dfrac{\partial \mu_{O}}{\partial b_{\text{sp}}}\dfrac{\partial \mu_{O}}{\partial b_{\text{ph}}} \\ [2em]
\dfrac{\partial \mu_{O}}{\partial b_{\text{sp}}}\dfrac{\partial \mu_{O}}{\partial b_{\text{ph}}} & \left(\dfrac{\partial \mu_{O}}{\partial b_{\text{ph}}}\right)^2 
\end{array}\right)=\dfrac{1}{\sigma_{O}^{2}}\left(\begin{array}{cc}
\dfrac{4 b_{\text{sp}}^{2}}{b_{\text{ph}}^{4}} & -\dfrac{4 b_{\text{sp}}^{3}}{b_{\text{ph}}^{5}} \\ [2em]
-\dfrac{4 b_{\text{sp}}^{3}}{b_{\text{ph}}^{5}} & \dfrac{4 b_{\text{sp}}^{4}}{b_{\text{ph}}^{6}} 
\end{array}\right).
\label{fisher_O}
\end{equation}
More explicitly, the $F^{O}_{11}$ element of the above matrix will be added to the $F_{ij}$ element of the total Fisher matrix corresponding to the spectroscopic galaxy bias ($\lambda_i=\lambda_j=b_{\text{sp}}$). The element $F^{O}_{22}$ will be added to the element corresponding the photometric galaxy bias ($\lambda_i=\lambda_j=b_{\text{ph}}$). Finally, the two remaining elements of this new Fisher matrix will be additional terms in the off-diagonal elements involving both the spectroscopic and photometric galaxy biases ($\{\lambda_i,\lambda_j\}=\{\lambda_j,\lambda_i\}=\{b_{\text{sp}},b_{\text{ph}}\}$). We will address in the next section how we can build from an existing  Fisher matrix for the main cosmological probes a new matrix with these additional terms. We denote this new way of combining the 2D and 3D probes as XC2, as opposed to XC that will be used to denote the baseline analysis GC$_{\rm sp}$+GC$_{\rm ph}$+WL+XC. We assess now the impact of the combination by examining its effects on the FoM.

\begin{figure}
    \begin{center}
    \includegraphics[scale=0.6]
    {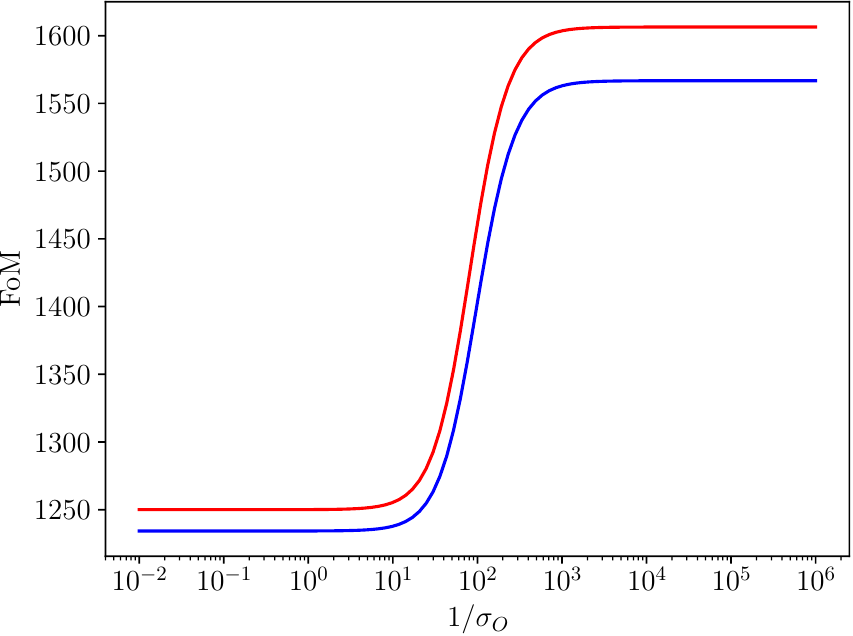}
    \caption{Evolution of the FoM as a function of the inverse of the standard deviation of the new observable, for the test cases considered in Sect.~\ref{discussion}. In blue, the case where fiducial values of the spectroscopic biases are taken to be equal to the photometric biases (bottom plateau at FoM=1234 and top plateau at FoM=1567).
In red, the case with different fiducial values of the spectroscopic biases and photometric biases {following} 
the baseline of \citetalias{IST:paper1} (bottom plateau at FoM=1250 and top plateau at FoM=1606). The FoM includes the information from GC$_{\rm sp}$, 3$\times$2\,pt, and the new observable.}
     
    \label{fig:flatness_common_bias}
    \end{center}
    \end{figure}
     
    \subsection{Discussion}\label{discussion}
    
    The objective of the present subsection is to provide an illustration of the benefits from the previously described approach. Firstly, in order to assess the gain obtained by introducing the new observable, we first omit it and compute a baseline FoM, based on the same ingredients and methodology as the optimistic scenario\,\footnote{More details on the settings describing the optimistic scenario are provided in  Sect. \ref{section:results}.} presented in~\citetalias{IST:paper1}. Our only difference is the use of a new redshift binning of the spectroscopic survey to allow {for the same effective redshifts} 
    for the spectroscopic and the photometric surveys in their overlapping range. The resulting FoM is calculated to be $1250$, close to the value of $1257$ obtained in~\citetalias{IST:paper1}. Such a minor difference can be explained by the different redshift binning.

    Then, to validate our pipeline, a test case is considered where the spectroscopic galaxy bias is set equal to the photometric galaxy bias ($b_{\text{sp}}=b_{\text{ph}}$). 
    
    Next, the new observable $O$ is introduced and its impact on the FoM is assessed. For simplicity, it is assumed that the standard deviation of the new observable, $\sigma_{O}$, remains constant across all redshift bins (this assumption will be relaxed for the final results in the next section). The value of the FoM with respect to the variation of $\sigma_{O}$ is computed and depicted in Fig.~\ref{fig:flatness_common_bias}. When $\sigma_O$ is large (equivalently, $1/\sigma_O$ is small), the FoM approaches an asymptotic value of $1234$. This value is slightly lower than the previously obtained $1250$ FoM because the spectroscopic bias has been adjusted to match the photometric bias, resulting in a reduction of constraining power from the spectroscopic sample. Conversely, for a small value of $\sigma_{O}$, the FoM reaches a higher asymptotic value of 1567, indicating that the inclusion of the new observable has a beneficial impact on the FoM. 
    This increase in the FoM for small $\sigma_{O}$ is not surprising, as in this scenario, the two tracers essentially share similar values of the galaxy bias with great precision, effectively reducing the total number of degrees of freedom in the problem (i.e., the number of free parameters). Indeed, this asymptotic value of $1567$ is  recovered numerically using the pipeline used in~\citetalias{IST:paper1} when assuming a common bias, which 
     yields $1567$ and thus validates our entire pipeline.

    Finally, we conducted tests in which the spectroscopic and photometric galaxy biases were set to their fiducial values as presented in~\citetalias{IST:paper1} ($b_{\text{sp}}>b_{\text{ph}}$). For scenarios with a large $\sigma_{O}$, we obtained an FoM of $1250$, which aligns perfectly with the previously obtained value when the new observable was not included. This result is expected since, for large $\sigma_O$, the new observable does not provide additional information. Conversely, for scenarios with a small $\sigma_{O}$, the FoM increased to $1606$, showing a similar level of improvement as observed in the previous case with a common bias. 
    
\section{Expected improvement on \Euclid constraints}\label{sec:6}

\begin{figure*}[ht]
\centering
\includegraphics[width=0.9\textwidth]
{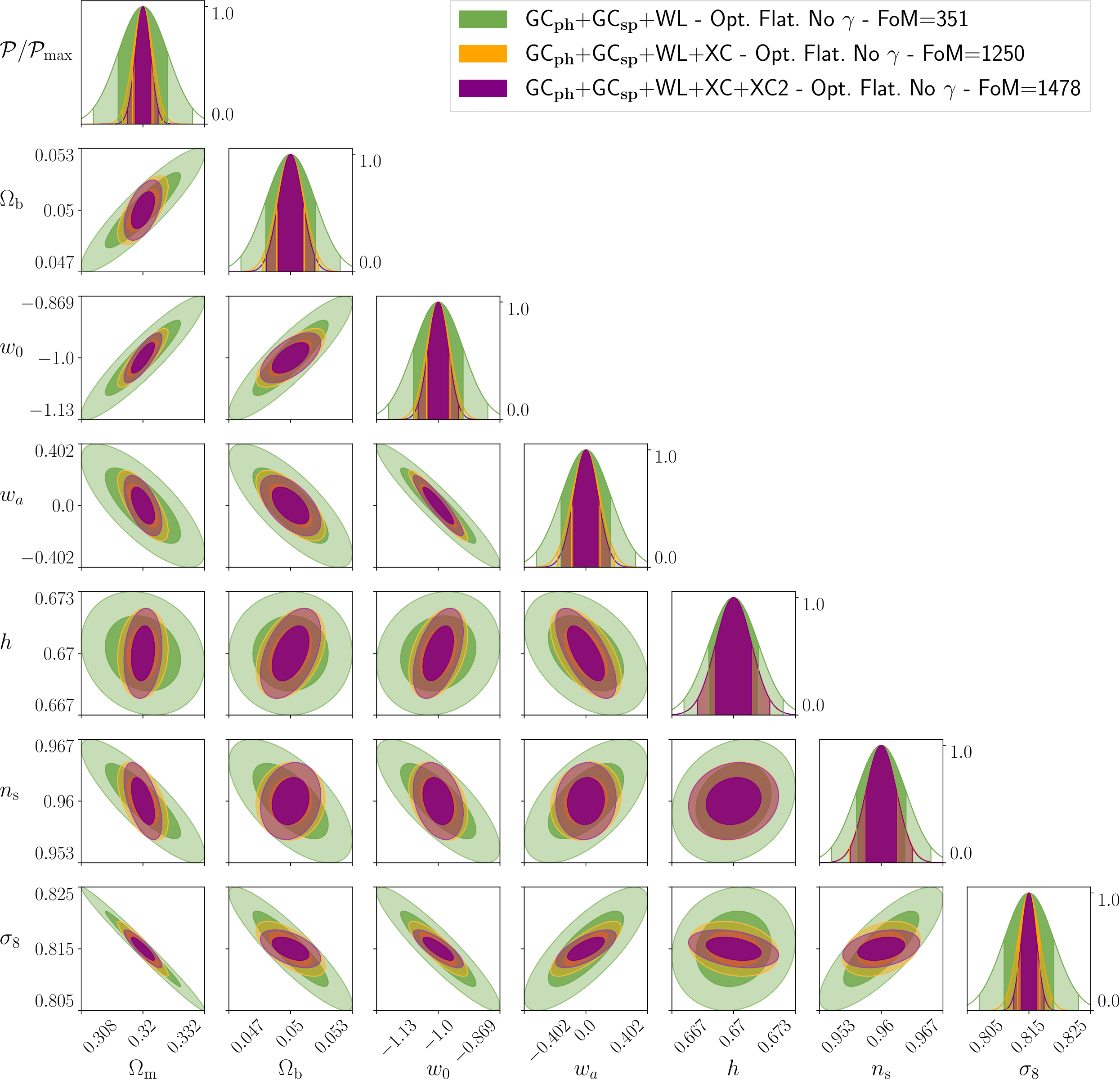}
\caption{1D normalized likelihood ($\mathcal{P} / \mathcal{P}_{\max }$) and 2D marginalised Fisher constraints (with 1$\sigma$ and 2$\sigma$ limits highlighted) for different combinations of probes. The most stringent constraints are obtained when including the new ratio observable (XC2). }
\label{triplot_GCp_WL}
\end{figure*}

    \subsection{Adjustment of the binning and galaxy selection function in the spectroscopic sample}\label{sec61}
    As seen in the previous section, a substantial  improvement in constraints can be achieved, provided the variance on the quantity $O$  is small enough. However, this new observable 
    requires that the selection functions for the two samples are identical (up to a normalisation factor).
     In order to forecast 
    the potential benefits of this method for the \Euclid constraints, 
    it is necessary to adapt the selection function of the spectroscopic galaxy sample to that of the photometric one,
     in order to  ensure that the former 
     is proportional  to the latter. 
    
    In practice, for each tomographic bin in the photometric sample, it is necessary to select the galaxies in the spectroscopic sample in such a way that the selection function for the two samples is identical. These selected galaxies will then constitute the corresponding population for the spectroscopic sample in the same redshift bin. Throughout this process, we make the assumption that the selection functions retain an identical form, thus ensuring consistency between the two samples. This objective could in principle be achieved by weighting the galaxies in a given sample by the inverse of the selection function, as long as the latter is known. However, the spatially inhomogeneous nature of the resulting Poisson noise could alter the effectiveness of the method. {We note that this new selection function for the spectroscopic sample is only considered for the new observable $O$, as detailed in Sect.\,\ref{sec:4}. When considering GC$_{\rm sp}$ we still consider narrow redshift bins,to benefit from the radial precision and include all the relevant effects in the modelling. The only difference for GC$_{\rm sp}$ compared to \citetalias{IST:paper1} is that we centre the narrow redshift bins on the effective redshifts of the photometric sample. This allows us to have essentially the same galaxy bias for the spectroscopic galaxies in GC$_{\rm sp}$ and the new observable.}

    Table~\ref{table:density2} presents the numerical values for the galaxy number density in each bin of the spectroscopic sample. The total density of all five bins adds up to 0.35 galaxies per square arcminute, which corresponds to the density of the entire spectroscopic sample. 
    {For the photometric probe, we consider the same binning presented in \citetalias{IST:paper1}, except for a minor change. Since the spectroscopic survey only goes {out} to redshift 1.8, we split the last tomographic bin of the photometric sample into two, with half its number density, and therefore consider 11 bins instead of 10. The binning and density values considered for the photometric sample in the redshift range where it overlaps with the spectroscopic sample is presented in Table~\ref{table:density3}.}

\begin{mytable}
    \begin{center}
    \caption{Expected number density of galaxies for the \Euclid spectroscopic survey per unit area and redshift intervals, $\mathrm{d}N / \mathrm{d}\Omega \, \mathrm{d} z\left[\mathrm{deg}^{-2}\right]$ for our new set of redshift bins (with respect to the baseline in \citetalias{IST:paper1}, Table 3) and the corresponding density of galaxies per arcmin$^{2}$ for each redshift bin ($n_{\rm gal}$).}
    \label{table:density2}
    \begingroup
    \setlength{\tabcolsep}{3pt} 
    \renewcommand{\arraystretch}{1.5} 
    \begin{tabular}{|c|c|c|c|c|c|}
    \hline
    {Redshift central bin} & 0.9595 & 1.087 & 1.2395 & 1.45 & 1.688 \\ \hline
    {$\mathrm{d}N / \mathrm{d}\Omega \, \mathrm{d} z\left[\mathrm{deg}^{-2}\right]$} & 1807.76 & 1793.63 & 1655.01 & 1320.51 & 870.13 \\ \hline
    {$\Delta z$ : width of bin} & 0.119 & 0.136 & 0.169 & 0.252 & 0.224 \\ \hline
    {$n_{\rm gal}$\,$\left[\text{ arcmin}^{-2}\right]$} & 0.0597 & 0.0677 & 0.0781 & 0.0924 & 0.0536 \\ \hline
    
    \end{tabular}
    \endgroup
    \end{center}
\end{mytable}

\begin{mytable}
    \begin{center}
    \caption{Expected number density of galaxies for the \Euclid photometric survey per unit area and redshift intervals, $\mathrm{d}N / \mathrm{d}\Omega \, \mathrm{d} z\left[\mathrm{sr}^{-1}\right]$ for our new set of redshift bins (with respect to the baseline in \citetalias{IST:paper1}, Table 4) and the corresponding density of galaxies per arcmin$^{2}$ for each redshift bin ($n_{\rm gal}$). {Only the redshift range that overlaps with the spectroscopic survey is shown.}}
    \label{table:density3}
    \begingroup
    \setlength{\tabcolsep}{2pt} 
    \renewcommand{\arraystretch}{1.5} %
\begin{tabular}{|c|c|c|c|c|c|}
\hline
{Redshift central bin} & 0.9595 & 1.087 & 1.2395 & 1.45 & 1.688 \\ \hline
{$\mathrm{d}N / \mathrm{d}\Omega \, \mathrm{d} z\left[\mathrm{sr}^{-1}\right]$} & 4219063 & 4821786 & 5991778 & 8934486 & 3970883 \\ \hline
{$\Delta z$ : width of bin} & 0.119 & 0.136 & 0.169 & 0.252 & 0.224 \\ \hline
{$n_{\rm gal}$,$\left[\text{ arcmin}^{-2}\right]$} & 3.0 & 3.0 & 3.0 & 3.0 & 1.5 \\ \hline
\end{tabular}

    \endgroup
    \end{center}
\end{mytable}

    \subsection{Final results}
    \label{section:results}
    \begin{figure*}[htbp]
    \centering
    \includegraphics[width=0.9\textwidth]{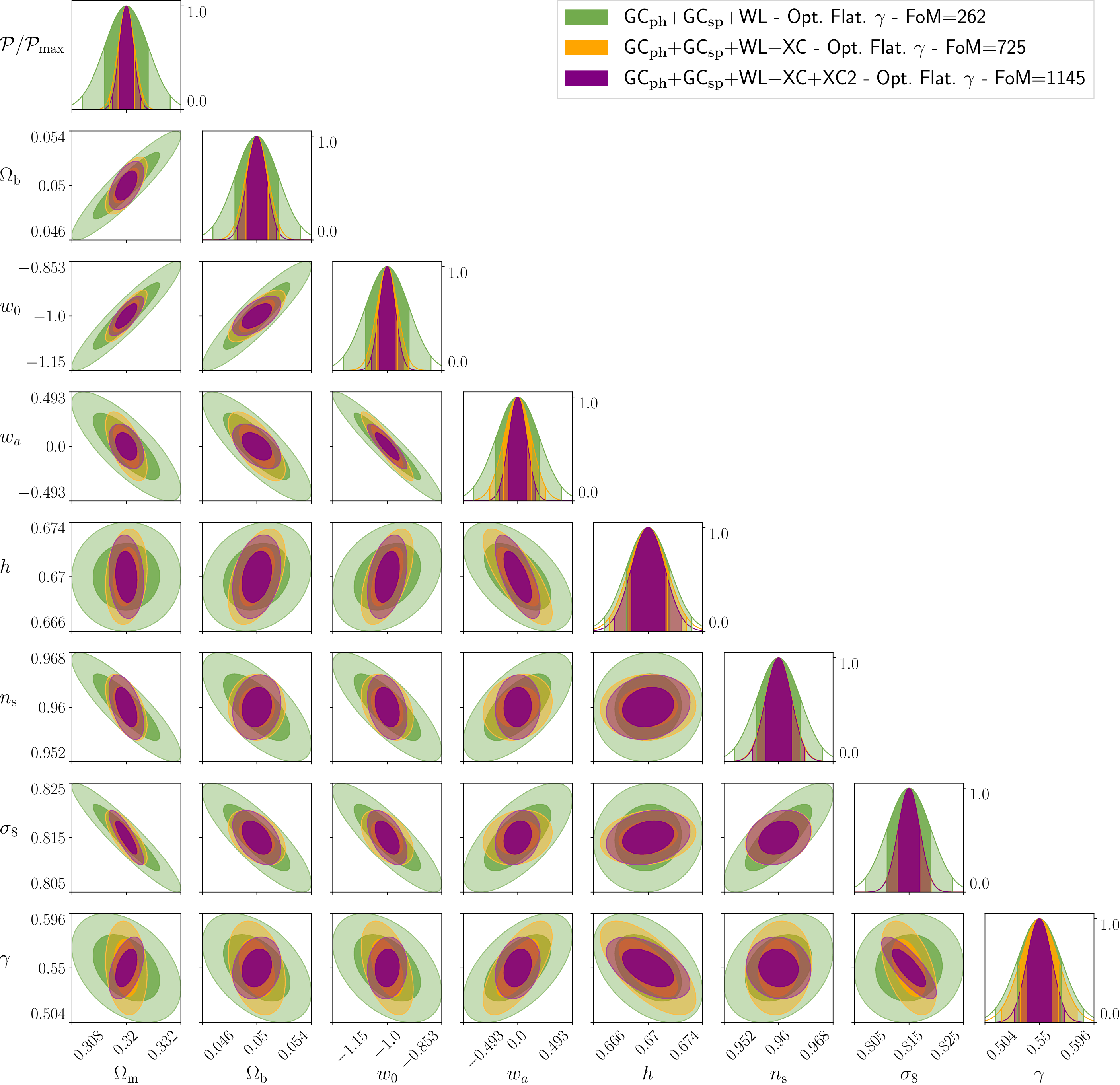}
    \caption{Same as in Fig.\,\ref{triplot_GCp_WL} for the modified gravity scenario, with the growth index $\gamma$ as additional parameter.
    }
    \label{triplot_GCp_WL_gamma}
    \end{figure*}
    
    For each redshift bin where data are available from both surveys, we estimate 
    the total Fisher matrix by adding the contribution from the new observable $O$ to the Fisher matrix computed in the standard case. This allows 
    us to monitor the improvement in various  \Euclid constraints coming from our new XC2 cross-correlation probe between photometric and spectroscopic surveys, compared 
    to the case where only the correlation between GC$_{\rm ph}$ and WL is taken into account (i.e., the 3$\times$ 2\,pt analysis). We consider 
    two cases in our comparison: the optimistic and the semi-pessimistic cases. In both cases, we start from $\ell_{\min} = 10$. For the former, in line with the optimistic case in \citetalias{IST:paper1}, we apply a cut in multipoles at $\ell_{\max}=5000$ for the weak lensing probe, a cut at $\ell_{\max}=3000$ for GC$_{\rm ph}$ and XC, and a cut at $k = 0.3 h$\,Mpc$^{-1}$ for the 3D clustering probe GC$_{\rm sp}$. For the semi-pessimistic case, the cut is applied at $\ell_{\max}=1500$ for WL, $\ell_{\max}=750$ for GC$_{\rm ph}$ and XC, and $k = 0.25 h$\,Mpc$^{-1}$ for GC$_{\rm sp}$. {We note that a linear galaxy bias model will probably break down at the very small scales probed in the optimistic case.  {If we were to}  consider a nonlinear galaxy bias model, the link between our new observable and the ratio of galaxy biases from Eq.\,(\ref{eq:mean value}) would change. {This might be possible to explore, but } the extension to higher-order galaxy bias models is beyond the scope of this work. We still provide the results for the semi-pessimistic scenario, for which a linear galaxy bias model will be more appropriate, and decide to show the results for the optimistic case to compare with the ones presented in \citetalias{IST:paper1}.}
    Figure~\ref{triplot_GCp_WL} shows the marginalised constraints for the cosmological parameters in the optimistic case of the $(w_0, w_a)$ scenario. The addition of the new observable 
    (XC2) clearly improves the constraints and modifies the orientation of correlation ellipses between some parameters, hinting at a breaking of degeneracies. Similar comparisons are performed and illustrated in Fig.~\ref{triplot_GCp_WL_gamma} for the case where the growth index $\gamma$ is left free. Although no specific behaviour emerges from this comparison, we notice that a degeneracy between $\sigma_8$ and $\gamma$ appears when adding the new observable XC2. The improvement on the individual cosmological parameters are summarised in Fig.~\ref{graph_all}. {This} provides a visual comparison of the relative errors on each cosmological parameter and the corresponding FoM for all the cases studied.  Although the XC2 method always leads to appreciable improvements, it seems that no regular behaviour can be identified, as some parameters are improved for some cases and not for others. 

\begin{table*}[ht!]
\setlength{\tabcolsep}{8pt} 
\renewcommand{\arraystretch}{1.3} 
\centering
\caption{FoM values with (XC2) and without (XC) the new observable $O$, and the corresponding relative improvement, for both the semi-pessimistic and optimistic scenarios, in flat and non-flat cosmologies, assuming general relativity. Improvements are less important when the cut on $\ell$ for the new observable is lower, but there is still a significant improvement, especially in the semi-pessimistic case.}

\label{tab:general_relativity_compact}
\begin{tabular}{|c|c|c|c|c|}
\hline
\multicolumn{5}{|c|}{General relativity} \\
\hline
\multirow{2}{*}{Cut on \( \ell \) for $O$} & \multirow{2}{*}{Scenario} & \multirow{2}{*}{Curvature} & \multirow{2}{*}{FoM - XC2/XC} & \multirow{2}{*}{Gain} \\
 &  &  &  &  \\ \hline
\multirow{2}{*}{$\ell=3000$} & \multirow{2}{*}{Optimistic} & Flat & $1477.5 / 1250$ & +18.2\% \\ \cline{3-5}
 &  & Non-Flat & $608.6 / 485$ & +25.5\% \\ \cline{2-5}
 \hline \multirow{2}{*}{$\ell=750$} & \multirow{2}{*}{Semi-Pessimistic} & Flat & $689.9 / 567.6$ & +21.5\% \\ \cline{3-5}
 &  & Non-Flat & $181.7 / 145.5$ & +24.9\% \\ \hline
\multirow{6}{*}{$\ell=300$} & \multirow{2}{*}{Optimistic} & Flat & $1413.4 / 1250$ & +13.1\% \\ \cline{3-5}
 &  & Non-Flat & $572.3 / 485$ & +18\% \\ \cline{2-5}
 & \multirow{2}{*}{Semi-Pessimistic} & Flat & $678.6 / 567.6$ & +19.6\% \\ \cline{3-5}
 &  & Non-Flat & $177.2 / 145.5$ & +21.8\% \\ \cline{2-5}

 & \multirow{2}{*}{Cut on all probes} & Flat & $423.1 / 241.7$ & +75.1\% \\ \cline{3-5}
 &  & Non-Flat & $151.1 / 93.7$ & +61.3\% \\ \hline

\multirow{6}{*}{$\ell=100$} & \multirow{2}{*}{Optimistic} & Flat & $1318.7 / 1250$ & +5.5\% \\ \cline{3-5}
 &  & Non-Flat & $521.1 / 485$ & +7.4\% \\ \cline{2-5}
 & \multirow{2}{*}{Semi-Pessimistic} & Flat & $637.2 / 567.6$ & +12.3\% \\ \cline{3-5}
 &  & Non-Flat & $164.1 / 145.5$ & +12.8\% \\ \cline{2-5}

 & \multirow{2}{*}{Cut on all probes} & Flat & $154.8 / 89.5$ & +73\% \\ \cline{3-5}
 &  & Non-Flat & $58.9 / 37.4$ & +57.5\% \\ \hline

\end{tabular}
\label{table_XC2_GR}
\end{table*}

\begin{table*}[ht!]
\setlength{\tabcolsep}{8pt} 
\renewcommand{\arraystretch}{1.3} 
\centering
\caption{FoM values with (XC2) and without (XC) the new observable $O$, and the corresponding relative improvement, for both the semi-pessimistic and optimistic scenarios, in flat and non-flat cosmologies, assuming modified gravity. Improvements are less important when the cut on $\ell$ for the new observable is lower, but there is still a significant improvement, especially in the semi-pessimistic case.}
\label{tab:modified_gravity_compact}
\begin{tabular}{|c|c|c|c|c|}
\hline
\multicolumn{5}{|c|}{Modified gravity} \\
\hline
\multirow{2}{*}{Cut on \( \ell \) for $O$}& \multirow{2}{*}{Scenario} & \multirow{2}{*}{Curvature} & \multirow{2}{*}{FoM - XC2/XC} & \multirow{2}{*}{Gain} \\
 &  &  &  &  \\ \hline
\multirow{2}{*}{$\ell=3000$} & \multirow{2}{*}{Optimistic} & Flat & $1144.5 / 725.1$ & +57.8\% \\ \cline{3-5}
 &  & Non-Flat & $565.2 / 464.4$ & +21.7\% \\ \cline{2-5}
  \hline \multirow{2}{*}{$\ell=750$} & \multirow{2}{*}{Semi-Pessimistic} & Flat & $516.5 / 334.2$ & +54.5\% \\ \cline{3-5}
 &  & Non-Flat & $171.4 / 140.9$ & +21.6\% \\ \hline
\multirow{6}{*}{$\ell=300$} & \multirow{2}{*}{Optimistic} & Flat & $1054.2 / 725.1$ & +45.4\% \\ \cline{3-5}
 &  & Non-Flat & $548.7 / 464.4$ & +18.2\% \\ \cline{2-5}
 
 & \multirow{2}{*}{Semi-Pessimistic} & Flat & $503.9 / 334.2$ & +50.8\% \\ \cline{3-5}
 &  & Non-Flat & $168.7 / 140.9$ & +19.7\%  \\ \cline{2-5}

& \multirow{2}{*}{Cut on all probes} & Flat & $332.9 / 183.6$ & +81.3\% \\ \cline{3-5}
 &  & Non-Flat & $127.6 / 92.6 $ & +37.8\% \\ \hline

\multirow{6}{*}{$\ell=100$} & \multirow{2}{*}{Optimistic} & Flat & $886.9 / 725.1$ & +22.3\% \\ \cline{3-5}
 &  & Non-Flat & $521 / 464.4$ & +12.2\% \\ \cline{2-5}
 & \multirow{2}{*}{Semi-Pessimistic} & Flat & $453.2 / 334.2$ & +35.6\% \\ \cline{3-5}
 &  & Non-Flat & $160.9 / 140.9$ & +14.2\% \\ \cline{2-5}

 & \multirow{2}{*}{Cut on all probes} & Flat & $ 139.9/ 87.85$ & +59.3\% \\ \cline{3-5}
 &  & Non-Flat & $45 / 33.2 $ & +35.5\% \\ \hline

\end{tabular}
\label{table_XC2_MG}
\end{table*}
    
    \begin{figure*}
     \includegraphics[width=0.95\textwidth,height=0.95\textheight,keepaspectratio]{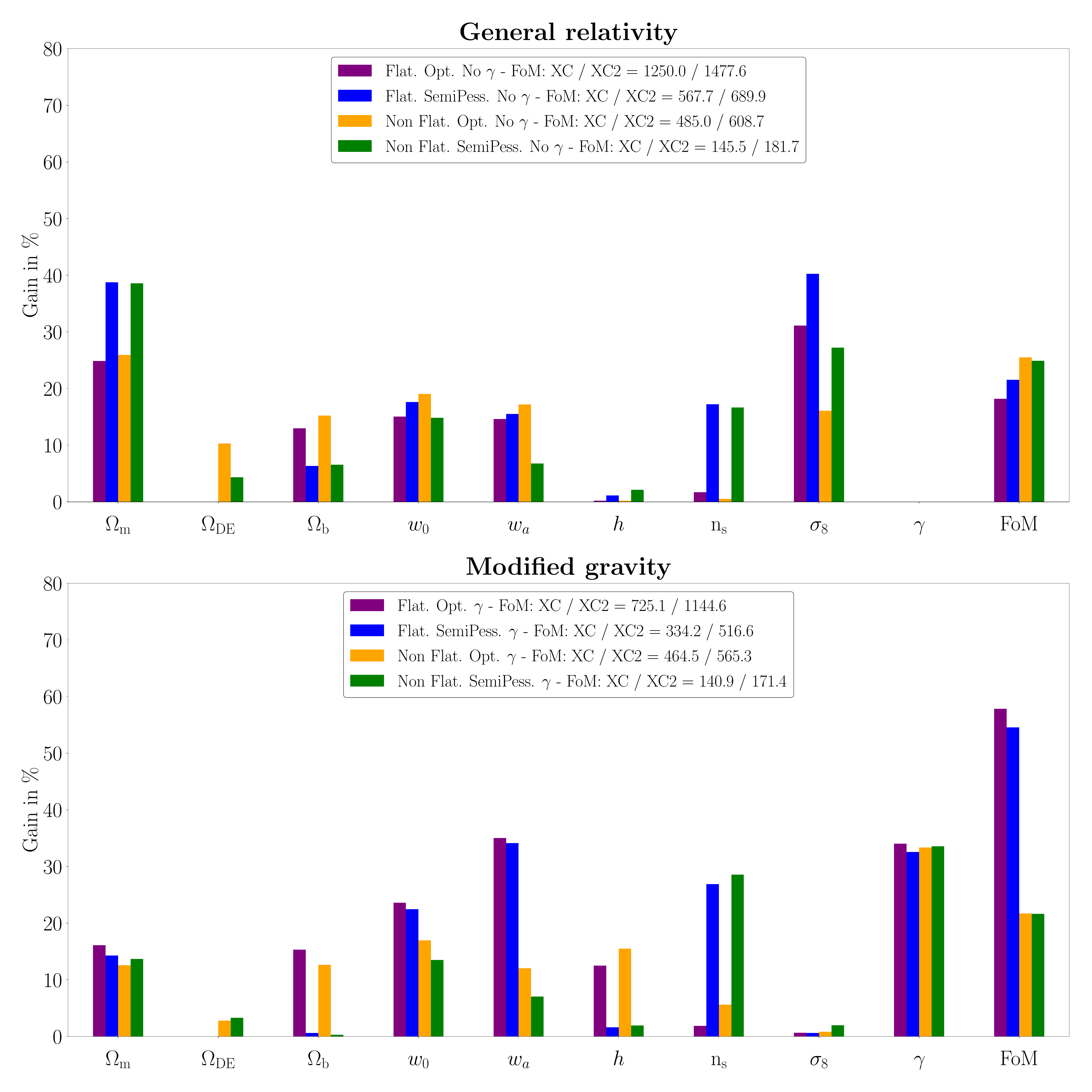}
      \caption{Improvement on cosmological constraints from adding the new observable XC2 compared to the baseline analysis. The top panel shows the improvement for the semi-pessimistic and optimistic settings, and flat and non-flat cosmologies within general relativity. The bottom panel shows the equivalent figure for the modified gravity scenario including the growth index $\gamma$. }
      \label{graph_all}  
      \end{figure*}
    
    The gains on the FoM are summarised in Table~\ref{table_XC2_GR} for the semi-pessimistic and
 optimistic scenarios, with a flat or
 non-flat cosmology, within general relativity. For the modified gravity case through the  $\gamma$ model, the comparison of FoMs is 
 presented  in Table~\ref{table_XC2_MG}. 
    The percentages correspond to the relative improvement on constraints provided by the XC2 method compared to the baseline analysis. 
In the case of general relativity, the FoM exhibits gains ranging from 18\% to 25\%. However, when considering cases with $\gamma$, substantial gains are observed for both the semi-pessimistic and optimistic scenarios in the flat case. The FoM increases by up to 54\% 
    for the semi-pessimistic 
    case and 57\% for the optimistic 
    case.  In contrast, for the non-flat case with $\gamma$, the increase is {only} 
    21\%.   
    Interestingly, the improvements on the FoM are highest for the flat {cosmological} case with the modified gravity $\gamma$ model. We also emphasise that a significant improvement is obtained on $\gamma$, by a factor {that} seems to be roughly the same regardless of the scenario considered. This suggests that the XC2 {probe} is well-suited to bring 
interesting constraints for modified gravity models.

\subsection{Taking into account  correlations}\label{correlation_section}

\begin{figure}
    \centering
 
    \includegraphics[scale=0.33]{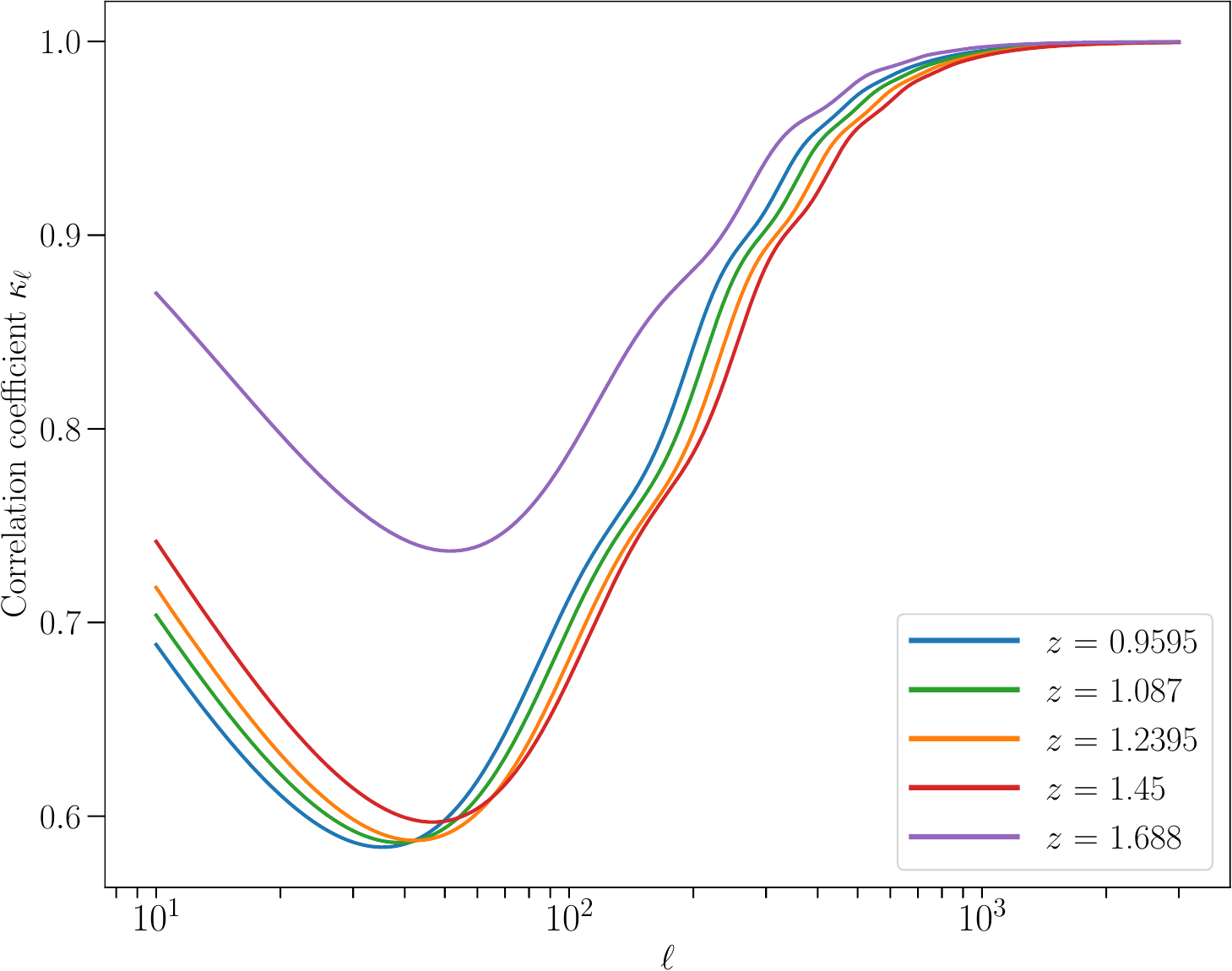}
    \caption{Correlation coefficient $\kappa_{\ell}$ as a function of $\ell$}
    \label{kappa_plot}
    \end{figure}

    The results of  previous sections were obtained assuming vanishing cross-correlations between $O$ and the $C_\ell$. Such terms would appear in the non-diagonal part of the covariance matrix. We have, therefore,  evaluated the correlation coefficient $\kappa_{\ell}$ between the two observables $o_{\ell}$ (Eq. \ref{eq:observable}) and $C_{\ell,\rm sp}$:
    \begin{equation}
    \kappa_{\ell}=\frac{\left\langle (\hat{o}_{\ell}-\bar{o}_{\ell})\left({\widehat{C_{\ell, \mathrm{sp}}}}-\overline{C_{\ell, \mathrm{sp}}}\right)\right\rangle}{\sigma_o\,\sigma_C},
    \label{kappa_ell}
    \end{equation}
    in which $\sigma_C^2$ stands for the variance of the $\widehat{C_{\ell,\rm sp}}$, which is given by
    \begin{equation}
    \sigma_C= \sqrt{\dfrac{2}{(2\ell+1) f_{\rm sky}}} \left( \overline{C_{\ell,\rm sp}}+\frac{1}{n_{\rm sp}}\right)\,.
    \end{equation}

The  correlation coefficient $\kappa_{\ell}$  has been estimated accordingly to Eq.\,\eqref{kappa_ell} {and is {plotted}} 
against $\ell$  
in Fig.\,\ref{kappa_plot}. 
Clearly the coefficient $\kappa_{\ell}$ is significant on all scales. The high correlation may undermine the effectiveness of the multi-tracer method as outlined in its simplified version above.   On the other hand, employing suitable weighting schemes may help mitigate this degradation \citep{2009PhRvL.103i1303S,2010PhRvD..82d3515H}, but this is beyond the scope of this paper.

\subsection{Further strategies for our multi-tracer method.}\label{fsky_section}

In order to see whether our method could still be useful, we have studied different strategies to get around the difficulty resulting from these strong correlations. 
In the first strategy, we assume that a  fraction of the survey is devoted {to} the standard combination of probes in our XC2  synthesis,  while  the other part $1-\alpha$ of the survey is used only for the determination of the observable $O$. In such a configuration, it is reasonable to assume that the volumes for each observable are independent {and therefore $O$ is not correlated to the other probes}. The FoM obtained with this strategy is represented by the blue curves in 
Figs.\,\ref{FoM_fsky_NO_GAMMA_plot} and \ref{FoM_fsky_GAMMA_plot}{, where we have considered the optimistic settings}. While in the standard scenario (Fig.\,\ref{FoM_fsky_NO_GAMMA_plot}), this strategy does not lead to any improvement in the FoM, one can see  from Fig.\,\ref{FoM_fsky_GAMMA_plot}  some gain can be obtained in the {modified gravity $\gamma$} 
model{. In more detail, by devoting 87\,\% of the sky coverage to the standard combination of probes and 13\,\% to the computation of the new observable,} the FoM is 9.4\,\% higher with our multi-tracer approach. This improvement is modest compared to the 58\% obtained in Table\,\ref{table_XC2_MG}, but {it} demonstrates that our method {can} 
still {provide} 
additional information.

In the second strategy, we assume that the 3$\times$2\,pt analysis is kept for the full survey, while only a fraction $\alpha$ is used for the $\GCsp$ analysis and $(1-\alpha)$ for the observable $O$. This case is represented by the orange curve {in Figs.\,\ref{FoM_fsky_NO_GAMMA_plot} and \ref{FoM_fsky_GAMMA_plot}}. In the standard model, a modest but non-zero  improvement can be seen (4.4\%) {when 71\,\% of the survey area is used for GC$_{\rm sp}$ and 29\,\% is used for the new observable}. This improvement is more appreciable in the the $\gamma$ model with a 30\,\% improvement for $\alpha = 0.57$. 

In the final strategy, we assume that the observable $O$ is derived from an independent survey. This assumption necessitates a thorough understanding and mastery of the (photometric) selection functions for {the} $\GCsp$ and $\GCph$ surveys. The use of an independent survey then becomes feasible for accessing the observable $O$. For instance, in the {Legacy Survey of Space and Time (LSST) of the Vera C. Rubin Observatory,} 
a significant portion of the observed sky will overlap with the \Euclid survey. 
This, in principle, should enable a precise understanding of the {(photometric)} selection functions employed in \Euclid  ($\GCsp$ and $\GCph$). Subsequently, the  portion of the LSST survey not covered by \Euclid can be utilized to estimate the observable $O$. This case is exemplified by the green curves in Figs.\,\ref{FoM_fsky_NO_GAMMA_plot} and \ref{FoM_fsky_GAMMA_plot}. As one can observe, a significant enhancement is achieved even for $\alpha \simeq$ 0.2$-$0.3. The improvement is consistently more pronounced in the case of the $\gamma$ model, suggesting that our multi-tracer approach yields a more substantial benefit in the context of modified gravity theories. However, it is important to note that the expected gain should be evaluated on a case-by-case basis. {We note that for this last strategy we consider values of $\alpha$ up to 1.2. This illustrates the improvement on the FoM when using an independent survey with a sky coverage even up to 20\,\% larger than \Euclid.}

\begin{figure}[htbp!]
    \centering

    \includegraphics[scale=0.62]{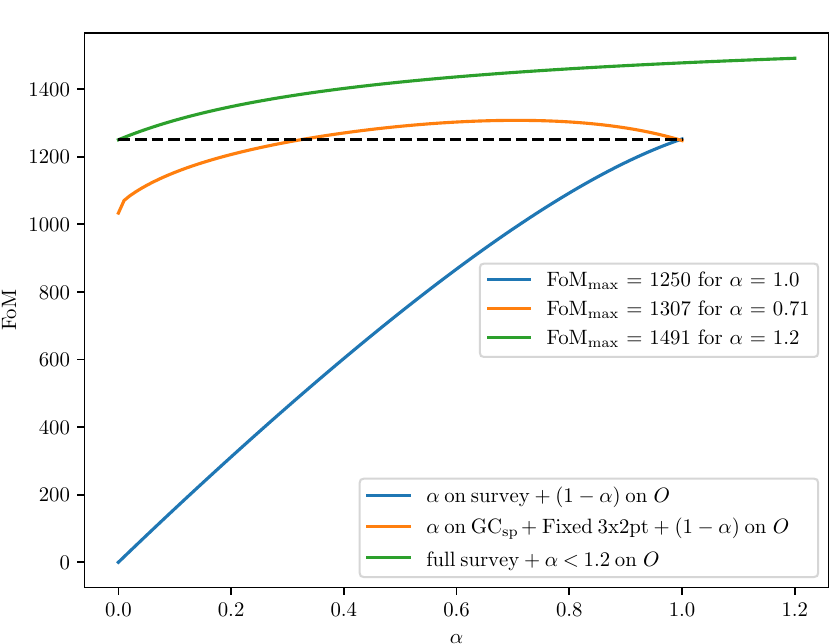}
 \caption{Figure-of-Merit (FoM) as a function of the fraction $\alpha$ of Euclid's observed sky ($f_{\text{sky}}$). The blue curve represents the combination of a fraction $\alpha$ for the main probes ($\GCsp$ and 3x2pt) and $(1-\alpha)$ for the new observable $O$. The orange curve represents the combination of a fraction of $\alpha$ for $\GCsp$, together with the full 3x2pt analysis for the full survey and a contribution of $(1-\alpha)$ for the new observable $O$. The green curve shows the FoM of the full survey with an additional external contribution from the new observable up to a factor $\alpha = 1.2$. The dashed line represents the optimal FoM using the standard $\GCsp$ and 3x2pt analysis. See the text for additional details on the different cases.} 
    \label{FoM_fsky_NO_GAMMA_plot}
\end{figure}

\begin{figure}[htbp!]
    \centering
    \includegraphics[scale=0.62]{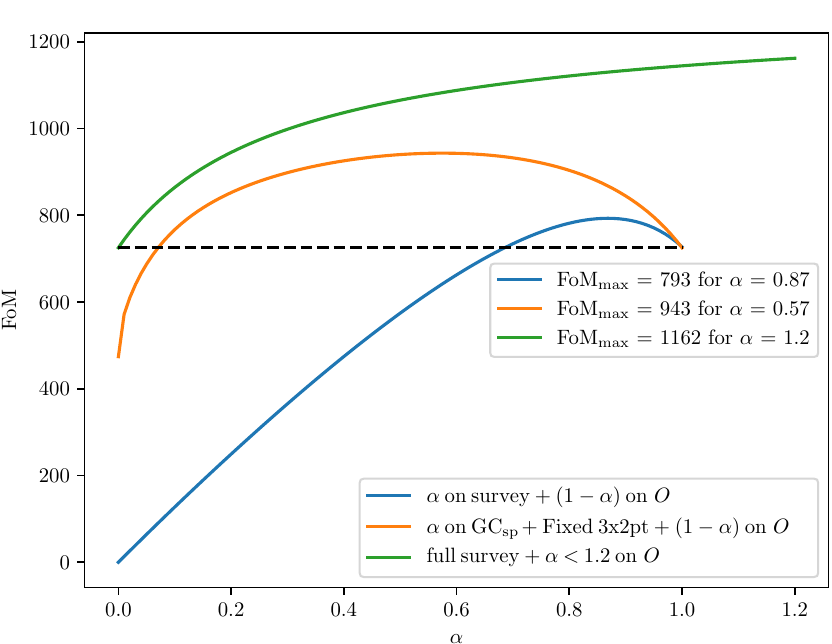}
    \caption{Figure-of-Merit (FoM) as a function of the fraction $\alpha$ of Euclid's observed sky ($f_{\text{sky}}$). We consider here the modified gravity model with the $\gamma$ parameter. The different curves correspond to the same cases treated in Fig\,\ref{FoM_fsky_NO_GAMMA_plot}.} 
    \label{FoM_fsky_GAMMA_plot}
    \end{figure}    

\section{Conclusions}\label{sec:7}
  
Spectroscopic galaxy surveys provide information on both the geometrical distribution of galaxies and their dynamics (through RSDs). Photometric galaxy surveys provide information on both the geometrical distribution of the tracers and the distribution of dark matter through its weak lensing imprint on the shape of galaxies. The information that can be inferred from a survey of a single tracer is primarily limited by the finite volume of the sample, i.e., the sampling noise. In the upcoming generation of wide-field surveys, there will be a significant overlap in the footprint of those two categories of survey. 
In the case of \Euclid, the {the area }overlap between the spectroscopic survey and the photometric survey will be nearly 100\%. This overlap offers the opportunity to bring additional information via cross-correlations. Euclid Collaboration:~Paganin et al. (in prep.) address the issue of the joint use of the 2D and 3D surveys in \Euclid, including their cross-correlations as an additional data vector. Their conclusion is that {the} covariance between 2D and 3D data can be safely neglected and the addition of the 2D$\times$3D data vector does not significantly change the final constraints. However, when two tracers are available over the same volume, one can also infer the ratio of the bias of the two populations without being limited by the sampling noise \citep{1988A&A...206L..11A, PhysRevLett.102.021302}. Given the large number density of objects {that will be} observed by the upcoming surveys, the Poisson noise limitation is expected to be very small. In this paper, we have introduced a new observable quantity, the ratio of angular \mbox{(cross-)correlation} functions, which provides an additional data vector, enabling a simple implementation of {the} multi-tracer technique. Using the specifications of \Euclid, we have shown that this additional observable provides useful information, resulting in improved estimations of cosmological parameters and, thereby, the FoM of dark energy. Depending on the {settings,} 
this improvement can {vary}  {from} modest (5\%) to more substantial (up to 60\%). It therefore appears to be a promising approach for enhancing the constraints from future joint analyses when two probes sample the same volume. Interestingly, a gain is achievable even when the biases of the two probes are identical. The details of this gain do not seem to follow a regular pattern. While it leads to a clear improvement in the FoM, constraints on certain cosmological parameters are contingent upon the specific cases studied. For instance, in the general relativity case, no significant enhancement is observed in the constraint on the Hubble parameter, whereas an improvement is observed within the $\gamma$ model. The constraints on the main targeted parameters, namely the FoM and constraint on $\gamma$, consistently exhibit improvement across the various cases investigated. We conclude that the actual benefit of this method needs to be assessed on a case-by-case basis. Finally, we note 
that the {new} additional observable we introduced is strongly correlated with the $\GCsp$ {harmonic power spectra.}  
{However, we have proposed different strategies, namely splitting the survey into two smaller surveys, or using an independent external survey, to circumvent such a large correlation. With these strategies we show that the multi-tracer technique proposed in this work can still provide additional valuable information from the combination of probes.} 
    
\begin{acknowledgements}

S.~Ili\'c  thanks the Centre national d’\'etudes spatiales (CNES) which supports his postdoctoral research contract. \AckEC.
\end{acknowledgements}

\bibliography{references}

\begin{appendix}
\onecolumn 
\section*{Appendix}\label{Append_label}
In this appendix, we provide detailed computations for determining the variance of the observable $\hat{o}_\ell$. We begin by defining the estimator $\hat{a}_{\ell m,\text{sp}}$ for the spectroscopic survey as $\hat{a}_{\ell m,\text{sp}}=b_{\text{sp}}a_{\ell m}^{\text{DM}}+a^p_{\ell m,\text{sp}}$, where $b_{\text{sp}}$ represents the spectroscopic bias, assumed to be a constant (i.e. no  stochasticity in the bias) $a_{\ell m}^{DM}$ denotes the contribution from the matter distribution, and finally $a^p_{\ell m,\text{sp}}$ represents the contribution from the Poisson noise. For simplicity, we initially consider a full sky scenario ($f_{\text{sky}}=1$). The Poisson noise is characterized by $\left<a^p_{\ell m,\text{sp}}\right>=0$ and $\left<\left|a^p_{\ell m,\text{sp}}\right|^2\right>=\frac{1}{n_{\text{sp}}}$. The estimator of the angular power spectrum is given by 
\begin{equation}
    \widehat{C}_{\ell,\text{sp}}=\frac{1}{2\ell+1}\sum_m\left|\hat{a}_{\ell m,\text{sp}}\right|^2=\frac{1}{2\ell+1}\sum_m\left(
    b_{\text{sp}}^2\left|a^{\text{DM}}_{\ell m}\right|^2+2b_{\text{sp}}\text{Re}(a^{\text{DM}}_{\ell m}a^{p\;*}_{\ell m})
    +\left|a^p_{\ell m}\right|^2\right)\,.
\end{equation}

The average over Poisson realization yields
\begin{equation}
    \left<\widehat{C}_{\ell,\text{sp}}\right>=\frac{1}{2\ell+1}\sum_mb_{\text{sp}}^2\left|a^{\text{DM}}_{\ell m}\right|^2+
    \frac{1}{n_{\text{sp}}}=\overline{C}_{\ell,\text{sp}}+\frac{1}{n_{\text{sp}}}\,.
\end{equation}
Here, the quantity $\overline{C}_{\ell,\text{sp}}$ is introduced. For the photometric survey, similar expressions can be formulated, with the Poisson noise being negligible due a significantly larger $n_{\text{ph}}$ compared to $n_{\text{sp}}$. Consequently, for the photometric angular power spectrum, we have
\begin{equation}
\left<\widehat{C}_{\ell,\text{ph}}\right>\simeq\overline{C}_{\ell,\text{ph}}=\frac{1}{2\ell+1}\sum_mb_{\text{ph}}^2\left|a^{\text{DM}}_{\ell m}\right|^2\,.
\end{equation} 

The observable $\hat{o}_\ell$ can then be expressed as 
\begin{equation}
    \hat{o}_\ell=\frac{\widehat{C}_{\ell,\text{sp}}}{\widehat{C}_{\ell,\text{ph}}}\simeq
    \frac{\widehat{C}_{\ell,\text{sp}}}{\overline{C}_{\ell,\text{ph}}}=\left(\frac{b_{\text{sp}}}{b_{\text{ph}}}\right)^2+
    \frac{1}{(2\ell+1)\overline{C}_{\ell,\text{ph}}}\sum_m\left(2b_{\text{sp}}\text{Re}(a^{\text{DM}}_{\ell m}a^{p\;*}_{\ell m})
    +\left|a^p_{\ell m}\right|^2\right)\,.
\end{equation}
Hence the mean value of this estimator is given by
\begin{equation}
    \left<\hat{o}_\ell\right>=\left(\frac{b_{\text{sp}}}{b_{\text{ph}}}\right)^2+\frac{1}{n_{\text{sp}}\overline{C}_{\ell,\text{ph}}}\,.
\end{equation}

The variance is then derived from $\sigma_o^2(\ell)=\left<(\hat{o}_\ell-\left<\hat{o}_\ell\right>)^2\right>$. Using the property $a^*_{\ell m}=(-1)^ma_{\ell,-m}$ and applying the Isserlis' theorem to compute $\left<\left|a^p_{\ell m}\right|^2\left|a^p_{\ell m'}\right|^2\right>$ (assuming Isserlis' theorem valid for the Poisson noise), straightforward calculations lead to
\begin{equation}
\sigma_o^2(\ell)= \frac{2}{(2\ell+1)f_{\rm sky}n_{\text{sp}}\overline{C}^2_{\ell,\text{ph}}}\left(
2 \overline{C}_{\ell,\rm sp}
+\frac{1}{n_{\rm sp}}\right)\,.\nonumber
\end{equation}
Here, the $f_{\text{sky}}$ factor has been incorporated, and one recovers expression Eq.~(\ref{eq:variance o_ell}).

\end{appendix}

\end{document}